\documentclass[10pt,a4paper]{article}
\usepackage[dvips]{color}
\usepackage{epsfig}
\usepackage{epstopdf}
\usepackage{amsmath}
\usepackage{graphicx}
\usepackage{amssymb,amsmath}
\usepackage{amsmath}
\usepackage{tabularx}

\begin{document}
\title{\bf On a phenomenology of the accelerated expansion with a varying ghost dark energy}
\author{{M. Zh. Khurshudyan$^{a}$\thanks{Email:khurshudyan@yandex.ru}~~~and A. N. Makarenko$^{b}$\thanks{Email:andre@tspu.edu.ru} }\\
$^{a}${\small {\em Institute for Physical Research, National Academy of Sciences, Ashtarak, Republic of Armenia}}\\
$^{b}${\small {\em Tomsk State Pedagogical University, Tomsk, Russian Federation}}\\
}\maketitle

\begin{abstract}
Subject of our study it is the accelerated expansion of the large scale universe, where a varying ghost dark energy can take the role of the dark energy. The model of the varying ghost dark energy considered in this work it is a phenomenological modification of the ghost dark energy. Recently, three other phenomenological models of the varying ghost dark energy have been suggested and the model considered in this work will complete the logical chain of considered modifications. The best fit of theoretical results to the luminosity distance, has been used to obtain preliminary constraints on the parameters of the models. This does help us to reduce amount of discussion. On the other hand, detailed comparison of theoretical results with observational data has been left as a subject of another discussion elsewhere. Moreover, a look to considered models via $Om$ and statefinder hierarchy analysis is presented and discussed for different forms of interaction between the varying ghost dark energy and cold dark matter.
\end{abstract}

\section{Introduction}\label{sec:INT}
Purpose of this work is to present new phenomenological modification of the ghost dark energy due to the energy density of dark matter. On the other hand, discuss appropriate consequences interesting from the point of view of cosmology. Particularly, to see the capacity of such phenomenological modification for the problem of the accelerated expansion of the low redshift universe~\cite{Riess}~-~\cite{Verde}. Recently three other models of the varying ghost dark energy have been considered and various cosmological scenarios able to explain the accelerated expansion of the low redshift universe have been studied~\cite{Martiros0}~-~\cite{Martiros0_3}. Moreover, in Ref.~\cite{Martiros0_1}~-~\cite{Martiros0_3} the authors have found, that in an appropriate radiation dominated universe, which evolves to recent large scale universe with a specific model of the varying ghost dark energy, massless particle creation is possible. This is a motivation behind our interest towards such modifications. The model of the varying ghost dark energy of this paper has the following energy density
\begin{equation}\label{eq:VGDE}
\rho_{de} = \alpha \rho_{de}^{m} H + \beta H^{2},
\end{equation}
where $\alpha$, $\beta$ and $m$ are constants and should be determined from observational data. The first constraints on these parameters, obviously, are such that $\rho_{de} > 0$. Considered model of dark energy, Eq.~(\ref{eq:VGDE}), compared to the ghost dark energy~($\rho_{de} = \alpha_{0} H + \beta H^{2}$), is modified admitting a function of the energy density of dark mater instead of the coefficient $\alpha_{0}$~(constant). Particularly, we consider power low function of the energy density of dark matter~(in case of two fluid darkness of low redshift universe). Due to the fact, that we have new phenomenological model of dark energy, first of all, we will study a cosmological scenario where dark energy and dark matter does not interact. On the other hand, to complete our study, we will consider various forms of interaction between those two components. We left full and detailed comparison of our theoretical results to observational data as a subject of another discussion elsewhere, but the best fit of theoretical results to the luminosity distance has been used to have preliminary constraints on the parameters of the models. This allows us to reduce amount of discussion and save a place. On the other hand, a look to considered models via $Om$~\cite{M0} and statefinder hierarchy analysis~\cite{M0_1} is performed and appropriate discussion is organized completing estimation of the present day values of $(r,s)$~\cite{M0_2} and $(\omega^{\prime}_{de}, \omega_{de})$~\cite{M0_3} for different values of interaction parameter $b$. The ghost dark energy it is one of the models of dark energy~\cite{Veneziano0}~-~\cite{Veneziano12}. There is an active discussion on different possibilities for dark energy models~\cite{Yoo}, various dark energy fluids including Chaplygin gas~(having various interesting modifications), varying cosmological constants, interacting dark energy models and viscous dark fluid models~\cite{M1}~-\cite{M4}~(and references therein). Moreover, there is an active discussion on the models admitting various parameterizations either of the EoS parameter or EoS itself~\cite{S0}~-\cite{S2}~(and references therein). The ghost dark energy model, Eq.~(\ref{eq:VGDE}), maybe also understood as generalized inhomogeneous fluids which have been introduced in Ref.~\cite{S0} and Ref.~\cite{S1}. We consider and complete our study involving interactions between dark components due to an active discussion on this topic in Literature, where an interaction indicates a transfer from one component to the other one. Among existing forms of interaction considered in Literature, we will consider specific forms of linear and linear sign changeable interactions to which we will come in the next section. There is an impressive amount of discussion of the role of these interactions in modern cosmology and presented citations in this work with references therein provide appropriate information on this topic. Dark energy it is one of the possibilities to explain the accelerated expansion of the large scale universe. Among the other ideas relevant to the solutions of the problems of modern cosmology, we can find within a modification of general relativity, which creates an effective description of dark energy. Moreover, it gives fundamental seeds for the origin of dark energy. To save our place we refer our readers to Ref~\cite{O1}~-\cite{O2}~(and references therein) for a comprehensive understanding of the role of the modification of general relativity in modern cosmology and physics.\\\\
The paper is organized as follows: In section~\ref{sec:CM} we will present detailed description of suggested cosmological models containing new varying ghost dark energy. The minimal model among considered models will be non interacting model. On the other hand, to have a comprehensive picture, we will consider interacting models with linear and linear sing changeable interactions considered in Literature very intensively. To simplify our discussion we have organized $4$ subsections covering cosmographic aspects of the models. Moreover, starting from the first law of thermodynamics we have obtained the dynamics of entropy for the varying ghost dark energy for each case. Redshift dependent graphical behavior of the cosmological parameters gives intuition about the impact of interaction parameter on the behavior of these parameters. Analysis presented in section~\ref{sec:CM} has been completed by an estimation of the present day values of $(r,s)$ and $(\omega^{\prime}_{de}, \omega_{de})$ parameters. On the other hand, in section~\ref{sec:PSA} we have organized a look to considered cosmological models via $Om$ and statefinder hierarchy analysis. Finally, discussion on obtained results and possible future extension of considered cosmological models are summarized in section~\ref{sec:Discussion}.

\section{Cosmography of the models}\label{sec:CM}
Two fluid approximation is used to describe the dynamics of the low redshift universe. One of the fluids is taken to be varying ghost dark energy, Eq.~(\ref{eq:VGDE}), while the second dark component it is nonrelativistic cold dark matter, which is pressurelles fluid. Two fluid approximation assumes that the energy density and pressure of the effective fluid are
\begin{equation}
\rho_{eff} = \rho_{de} + \rho_{dm},
\end{equation}
\begin{equation}
P_{eff} = P_{de} + P_{dm},
\end{equation}
describing the Hubble parameter as
\begin{equation}
H^{2} = \frac{\rho_{eff}}{3},
\end{equation} 
for $8 \pi G = c = 1$ units. On the other hand, the dynamics of the energy densities of each dark component is governed according to
\begin{equation}\label{eq:rhoD}
\dot{\rho}_{de} + 3 H (\rho_{de} + P_{de}) = -Q, 
\end{equation}
\begin{equation}\label{eq:rhoM}
\dot{\rho}_{dm} + 3 H \rho_{dm}  = Q. 
\end{equation}
Having the form of the interaction term $Q$ will allow us to analysis appropriate cosmological scenarios. To simplify our discussion we organise appropriate subsections. Presented discussion is due to observational constraints on the models due to the SNeIa test, which is based on the distance modulus $\mu$ related to the luminosity distance $D_{L}$ by
\begin{equation}
\mu = m - M = 5log_{10}D_{L},
\end{equation}
where $D_{L}$ is defined as
\begin{equation}
D_{L} = (1+z) \frac{c}{H_{0}}\int_{0}^{z}\frac{dz^{\prime}}{\sqrt{H(z^{\prime})}}.
\end{equation}
The quantities $m$ and $M$ denote the apparent and the absolute magnitudes, respectively. There are many different SNeIa data sets, obtained with different techniques. In some cases, these different samples may give very different results. 

\subsection{Non interacting model}\label{ssec:NINT}
\begin{figure}[h!]
 \begin{center}$
 \begin{array}{cccc}
 \includegraphics[width=80 mm]{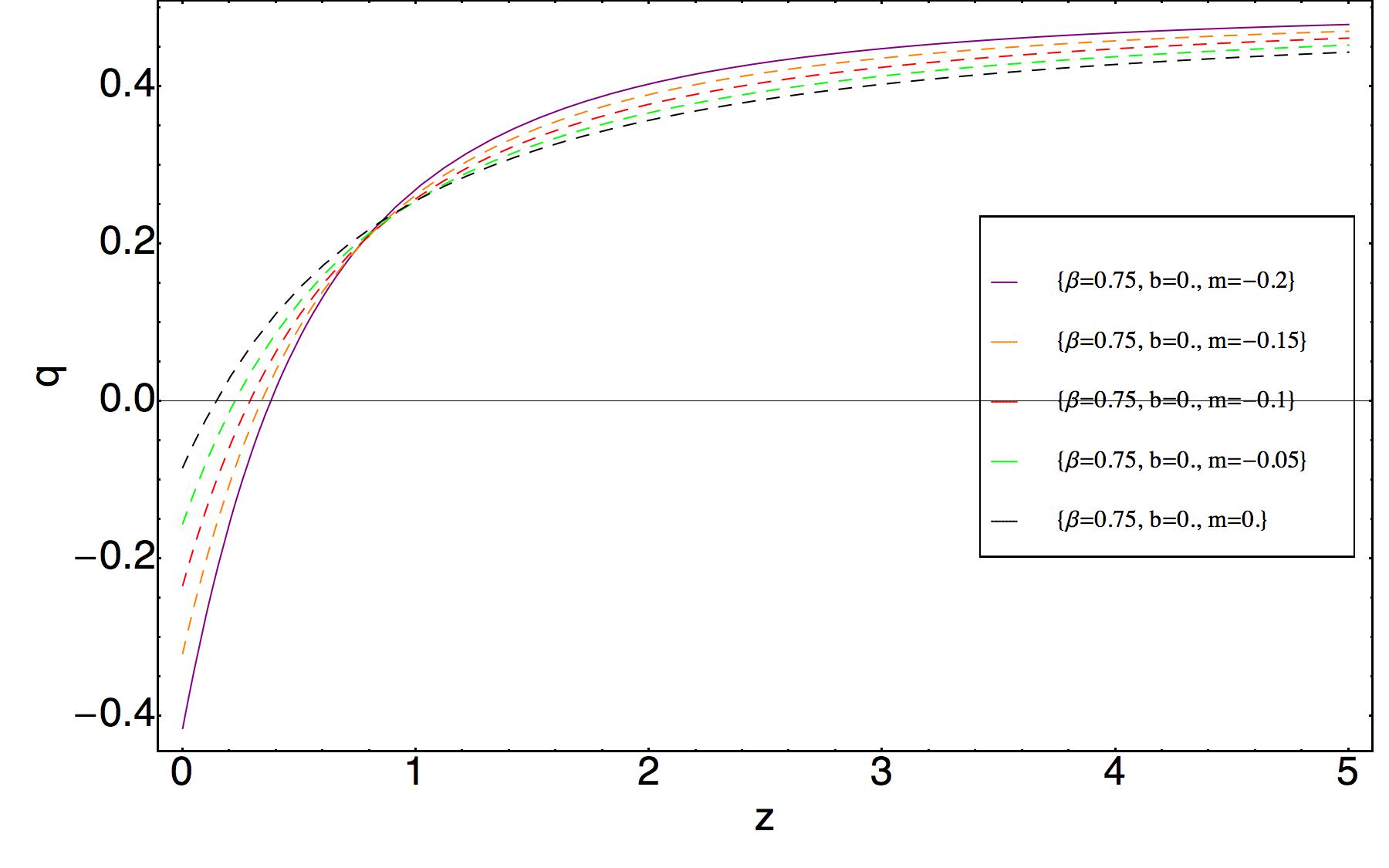}  &
\includegraphics[width=80 mm]{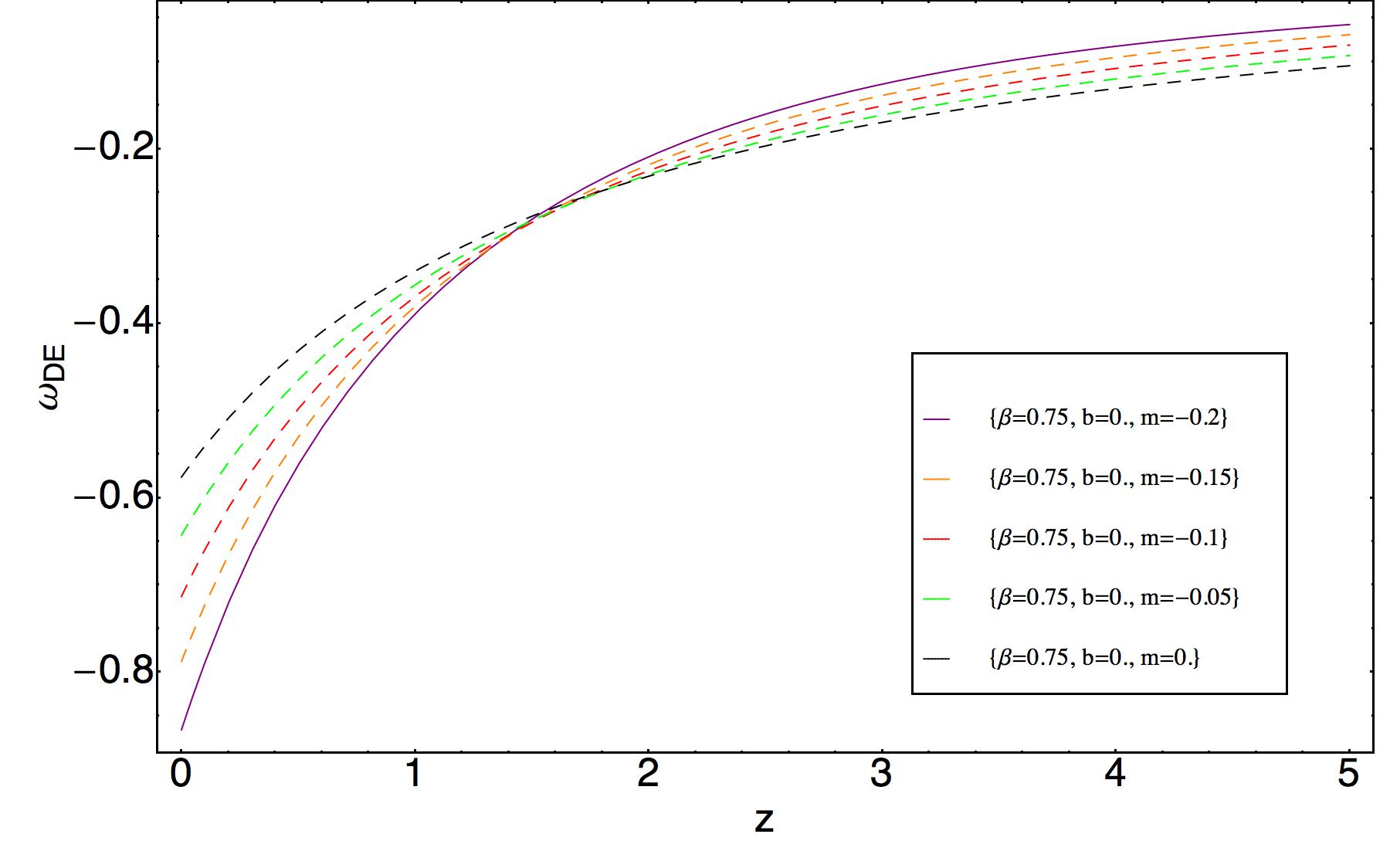}  \\
 \end{array}$
 \end{center}
\caption{Graphical behavior of the deceleration parameter $q$ and $\omega_{de}$ of non interacting varying ghost dark energy Eq.~(\ref{eq:VGDE}) against redshift $z$. $m=0$ does correspond to usual ghost dark energy.}
 \label{fig:Fig0}
\end{figure}

\begin{figure}[h!]
 \begin{center}$
 \begin{array}{cccc}
 \includegraphics[width=80 mm]{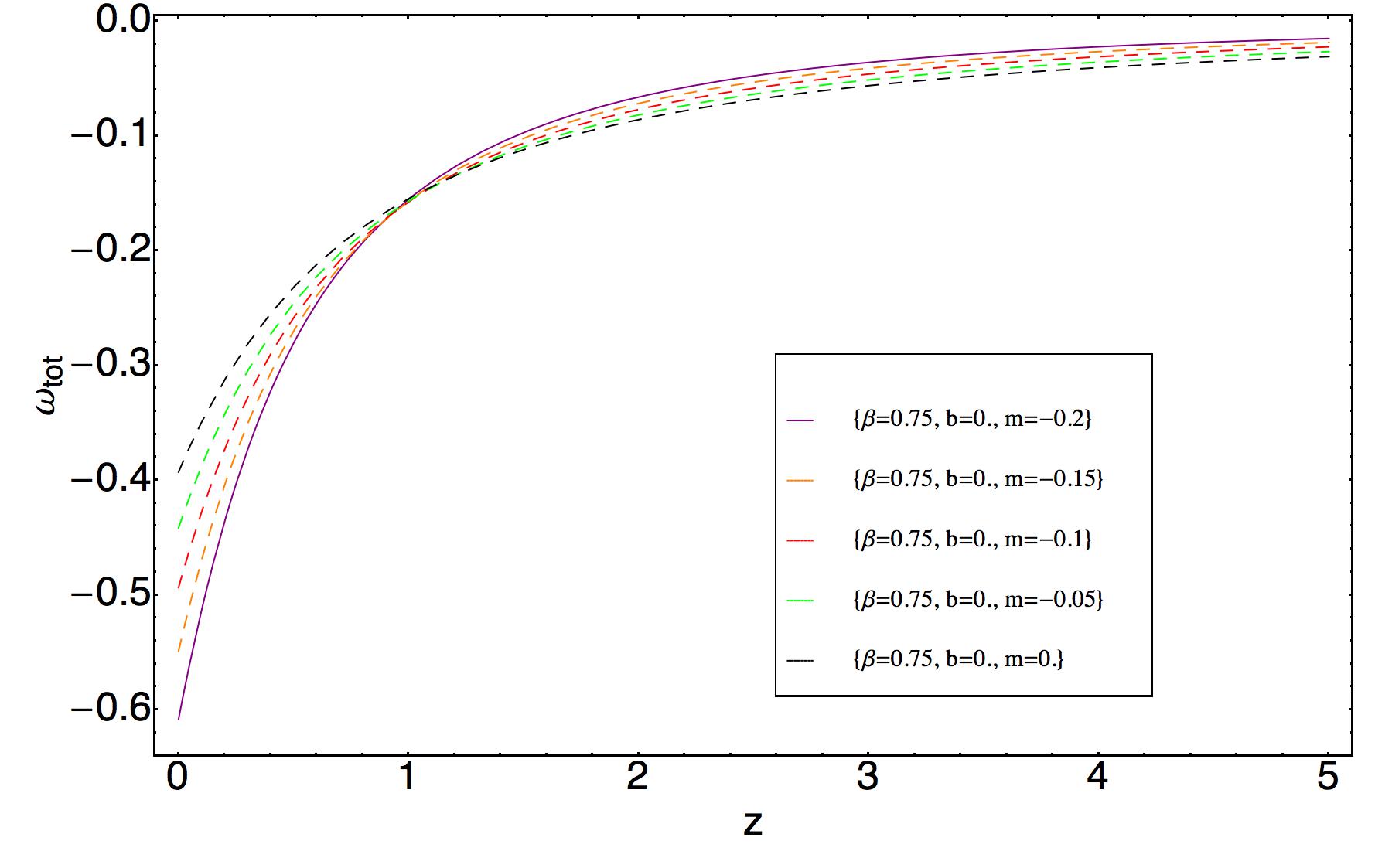}  &
\includegraphics[width=80 mm]{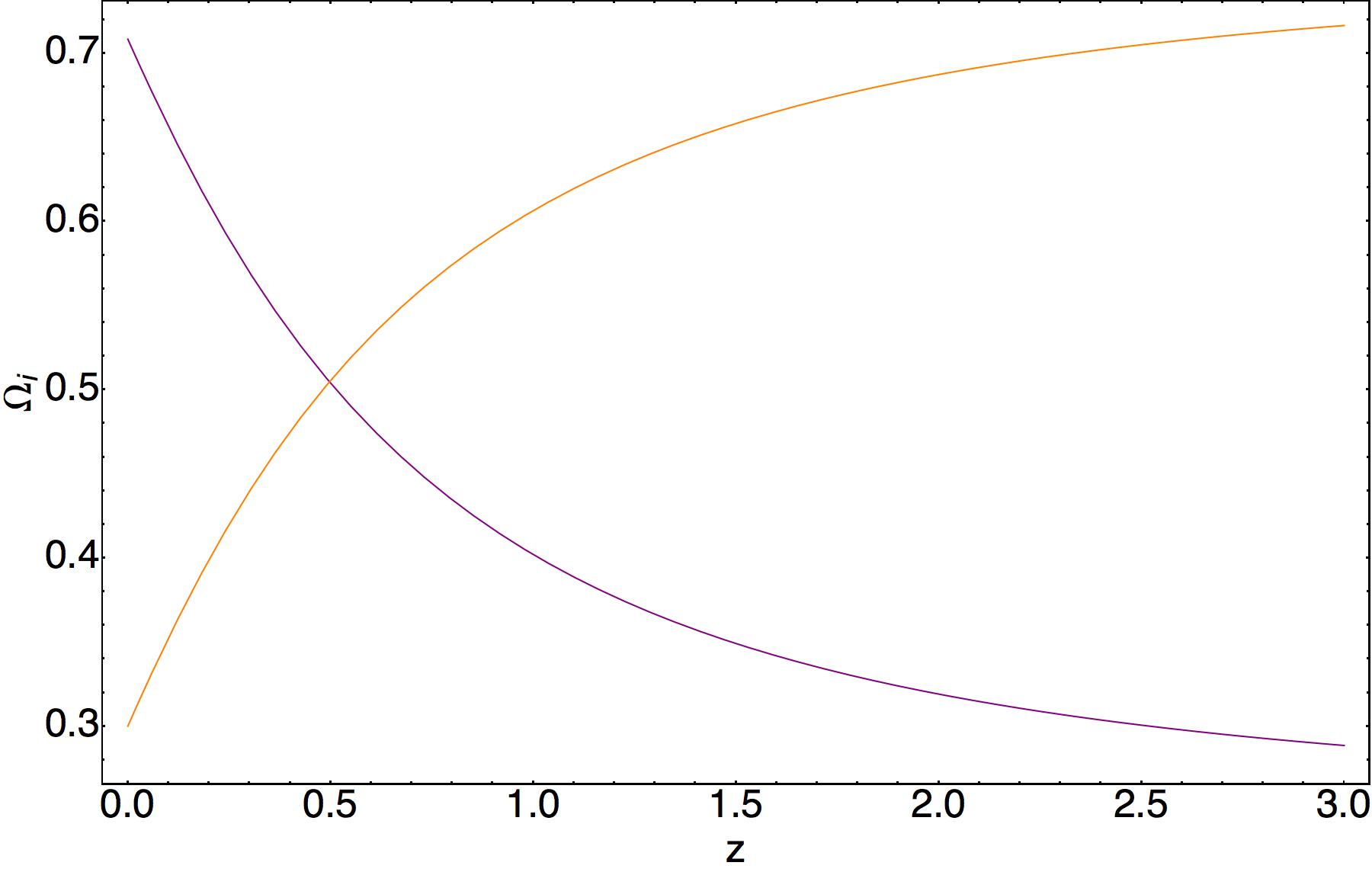}  \\
 \end{array}$
 \end{center}
\caption{Graphical behavior of the EoS parameter $\omega_{tot}$ of the effective fluid and $\Omega_{i}$ against redshift $z$. Considered behavior does correspond to non interacting varying ghost dark energy Eq.~(\ref{eq:VGDE}). $m=0$ does correspond to usual ghost dark energy. Behavior of $\Omega_{de}$ is represented by a blue curve on $\Omega_{i} - z$ plane, while orange line does represent behavior of $\Omega_{dm}$ (for $m=-0.2$).}
 \label{fig:Fig0_1}
\end{figure}

In modern cosmology non interacting dark energy model it is the minimal model compared to interacting models and should be studied first. According to Eq.~(\ref{eq:rhoD}) and Eq.~(\ref{eq:rhoM}), a non interacting model corresponds to $Q=0$ and in our case describes by the EoS of varying ghost dark energy of the following form
\begin{equation}\label{eq:omegadeNINT}
\omega_{de} = \frac{(2 m-1) \left(\beta -3 \Omega _{\text{de}}\right)}{\Omega _{\text{de}} \left(\beta +3 \Omega _{\text{de}}-6\right)}.
\end{equation}
It is easy to see that for the universe with $\Omega_{de} \to 0$~(probably a correct regime corresponding to very early universe) $\omega_{de} \to \infty$, while in case $\Omega_{de} = 1$ we have $\omega_{de} = 2 m-1$ asymptotic behavior. On the other hand, the deceleration parameter $q$ according to the form of $\omega_{de}$, Eq.~(\ref{eq:omegadeNINT}), accepts the following mathematical form
\begin{equation}\label{eq:qNINT}
q = \frac{(6-9 m) \Omega _{\text{de}}+\beta  (3 m-1)-3}{\beta +3 \Omega _{\text{de}}-6}.
\end{equation}
Moreover, despite to $\omega_{de} \to \infty$, the deceleration parameter $q$ is well defined function of $\beta$ and $m$ parameters
\begin{equation}
q = \frac{\beta (3 m-1)-3}{\beta - 6},
\end{equation}
while for $\Omega_{de} = 1$ regime
\begin{equation}
q = \frac{(\beta-3)(3m - 1) - 3}{\beta-3}. 
\end{equation}
Demand on $q \in [-1,0)$, $\omega_{de} \in [-1,0)$ and for simplicity $0 \leq \beta <1$, for the large scale universe gives the following constraints on $\Omega_{de}$ and new parameter $m$ 
\begin{equation}
\frac{1}{3}<\Omega _{\text{de}}\leq 1,
\end{equation}
\begin{equation}
\frac{\beta -\beta  \Omega _{\text{de}}-3 \Omega _{\text{de}}^2+3 \Omega _{\text{de}}}{2 \beta -6 \Omega _{\text{de}}}\leq m<\frac{\beta -6 \Omega _{\text{de}}+3}{3 \beta -9 \Omega _{\text{de}}}.
\end{equation}
On the other hand, considered model it is a cosmological model where the dynamics of the EoS parameter $\omega_{de}$ and $\Omega_{de}$ after some mathematics can be expressed in terms of $\Omega_{de}$, $\beta$ and $m$ as follows
\begin{equation}
\frac{d\omega_{de}}{dN} = -\frac{3 (1-2 m)^2 \left(\Omega _{\text{de}}-1\right) \left(\beta -3 \Omega _{\text{de}}\right) \left((\beta -6) \beta +6 \beta  \Omega _{\text{de}}-9 \Omega _{\text{de}}^2\right)}{\Omega _{\text{de}}^2 \left(\beta +3 \Omega _{\text{de}}-6\right){}^3},
\end{equation}
\begin{equation}
\frac{d\Omega_{de}}{dN} = \frac{3 (2 m-1) \left(\Omega _{\text{de}}-1\right) \left(\beta -3 \Omega _{\text{de}}\right)}{\beta +3 \Omega _{\text{de}}-6}.
\end{equation}
In Fig.~(\ref{fig:Fig0}) and Fig.~(\ref{fig:Fig0_1}) we have summarize redshift dependent graphical behavior of the deceleration parameter $q$, $\omega_{de}$, $\omega_{tot}$ and $\Omega_{i}$, respectively, indicating how the new parameter resulting from the phenomenological modification of ghost dark energy, Eq.~(\ref{eq:VGDE}), affects on the behavior of these parameters. From the graphical behavior presented in Fig.~(\ref{fig:Fig0}) and Fig.~(\ref{fig:Fig0_1}) we see, that at high redshifts  an increase of the value of $m$ parameter will decrease the deceleration parameter, EoS parameter of varying ghost dark energy and the EoS of the effective fluid. While, the increase of $m$ parameter will increase the values of the same parameters at low redshifts. Presented picture for the behavior of $\Omega_{de}$ and $\Omega_{dm}$~(Fig.~(\ref{fig:Fig0_1})) is according to the constraints on the parameters of the model for the best fit of theoretical results with distance modulus imposing $\beta=0.75$ and $m=-0.2$. Analysis of the graphical profile of the deceleration parameter shows that a phase transition from a decelerated expanding universe to the recent accelerated expanding universe is possible due to right dynamics of $\omega_{de}$, $\omega_{tot}$ and $\Omega_{de}$. Moreover, an increase of $m$ parameters is decreasing the transition redshift. In Fig.~(\ref{fig:Fig0}) and Fig.~(\ref{fig:Fig0_1}), reader can find an appropriate information indicating differences between cosmological models containing the ghost dark energy~($m=0$) and the varying ghost dark energy, respectively. One of the possibilities to study the dark energy model it is a reconstruction of its thermodynamics. For our model, when there is no interaction, after some trivial mathematics we can reach to the following differential equation describing the dynamics of entropy $S_{\text{de}}$   
\begin{equation}
T \frac{dS_{\text{de}}}{dH} = -\frac{4 \pi  \left((6-9 m) \Omega _{\text{de}}+\beta  (3 m-1)-3\right) \left(\Omega _{\text{de}} \left(\beta +3 \Omega _{\text{de}}-6 m-3\right)+\beta  (2 m-1)\right)}{H^2 \left(\beta +3 \Omega _{\text{de}}-6\right) \left(-3 (m-1) \Omega _{\text{de}}+\beta  m-3\right)},
\end{equation}
where $T$ it is the temperature. Statefinder analysis it is one of the tools developed to distinguish dark energy models. This is an analysis requiring to study two parameters in order to distinguish the models, and these parameters for our model are
\begin{equation}
r = 1 + \frac{9 (\omega_{\text{de}}  (\omega_{\text{de}} +1) \Omega _{\text{de}})}{2} + \frac{9 \left(\Omega _{\text{de}}-1\right) (1-2 m)^2 \left(\beta -3 \Omega _{\text{de}}\right) \left((\beta -6) \beta +6 \beta  \Omega _{\text{de}}-9 \Omega _{\text{de}}^2\right)}{2 \Omega _{\text{de}} \left(\beta +3 \Omega _{\text{de}}-6\right){}^3}
\end{equation}
\begin{equation}
s = \frac{2 \left(3 \Omega _{\text{de}} \left(-3 (m-1) \Omega _{\text{de}}+2 \beta  m-6\right)+((\beta -12) \beta +18) m+9\right)}{\left(\beta +3 \Omega _{\text{de}}-6\right){}^2}.
\end{equation}
In next three subsections we will consider interacting varying ghost dark energy models and discuss appropriate cases interesting for cosmological applications. We will start our analysis from the models where a classical form of interaction discussed in many papers will be taken into account, namely we will consider the following interaction
\begin{equation}
Q  = 3 b H (\rho_{de}+ \rho_{dm}),
\end{equation}
then we will analyze the models with sign changeable interactions between varying ghost dark energy and cold dark matter. 

\subsection{Interacting model $1$}\label{ssec:INT1}
\begin{figure}[h!]
 \begin{center}$
 \begin{array}{cccc}
 \includegraphics[width=80 mm]{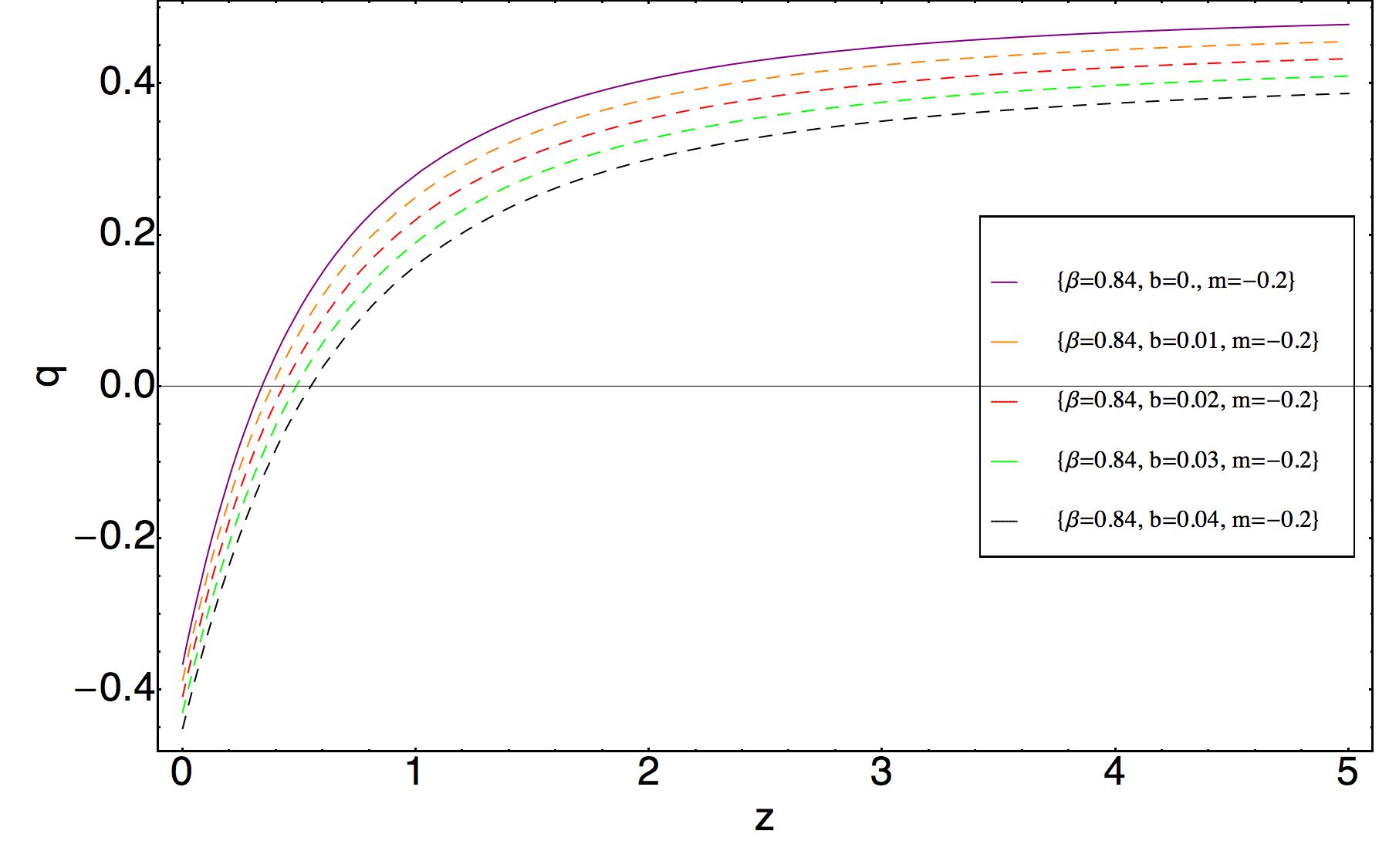}  &
\includegraphics[width=80 mm]{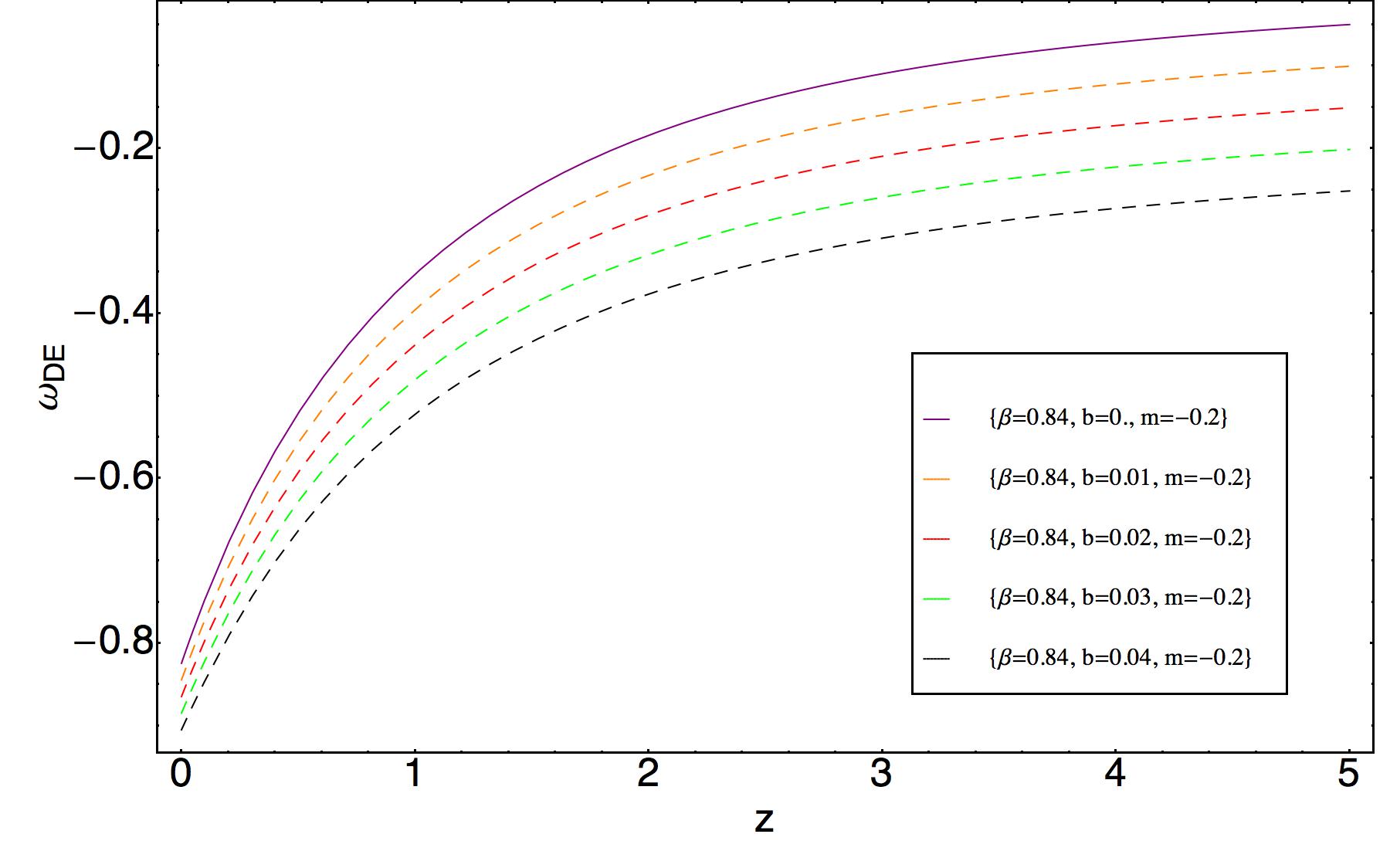}  \\
 \end{array}$
 \end{center}
\caption{Graphical behavior of the deceleration parameter $q$ and $\omega_{de}$ of interacting varying ghost dark energy Eq.~(\ref{eq:VGDE}) against redshift $z$. $m=0$ does correspond to usual ghost dark energy. The interaction is given via Eq.~(\ref{eq:INT1}).}
 \label{fig:Fig1}
\end{figure}

\begin{figure}[h!]
 \begin{center}$
 \begin{array}{cccc}
 \includegraphics[width=80 mm]{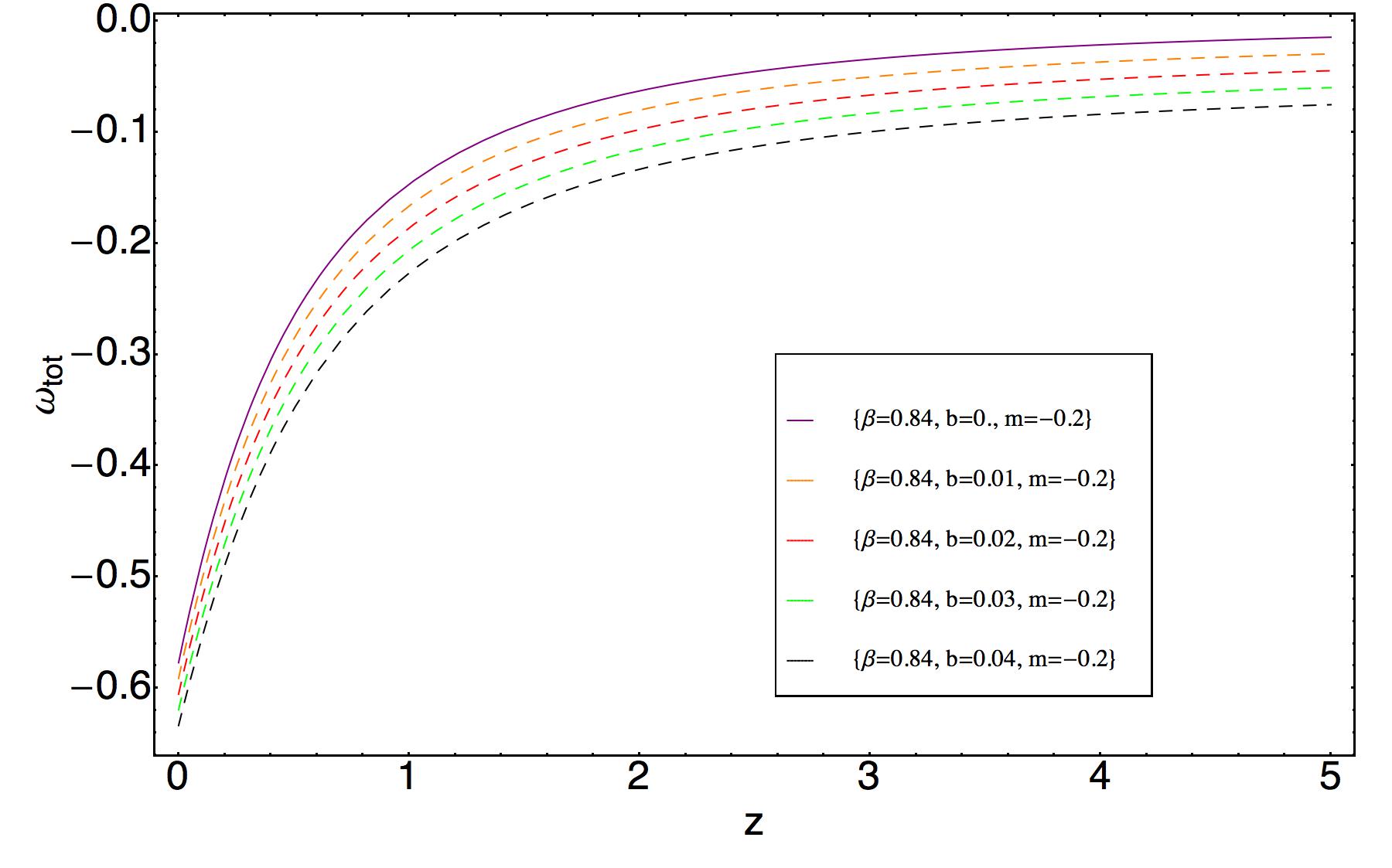}  &
\includegraphics[width=80 mm]{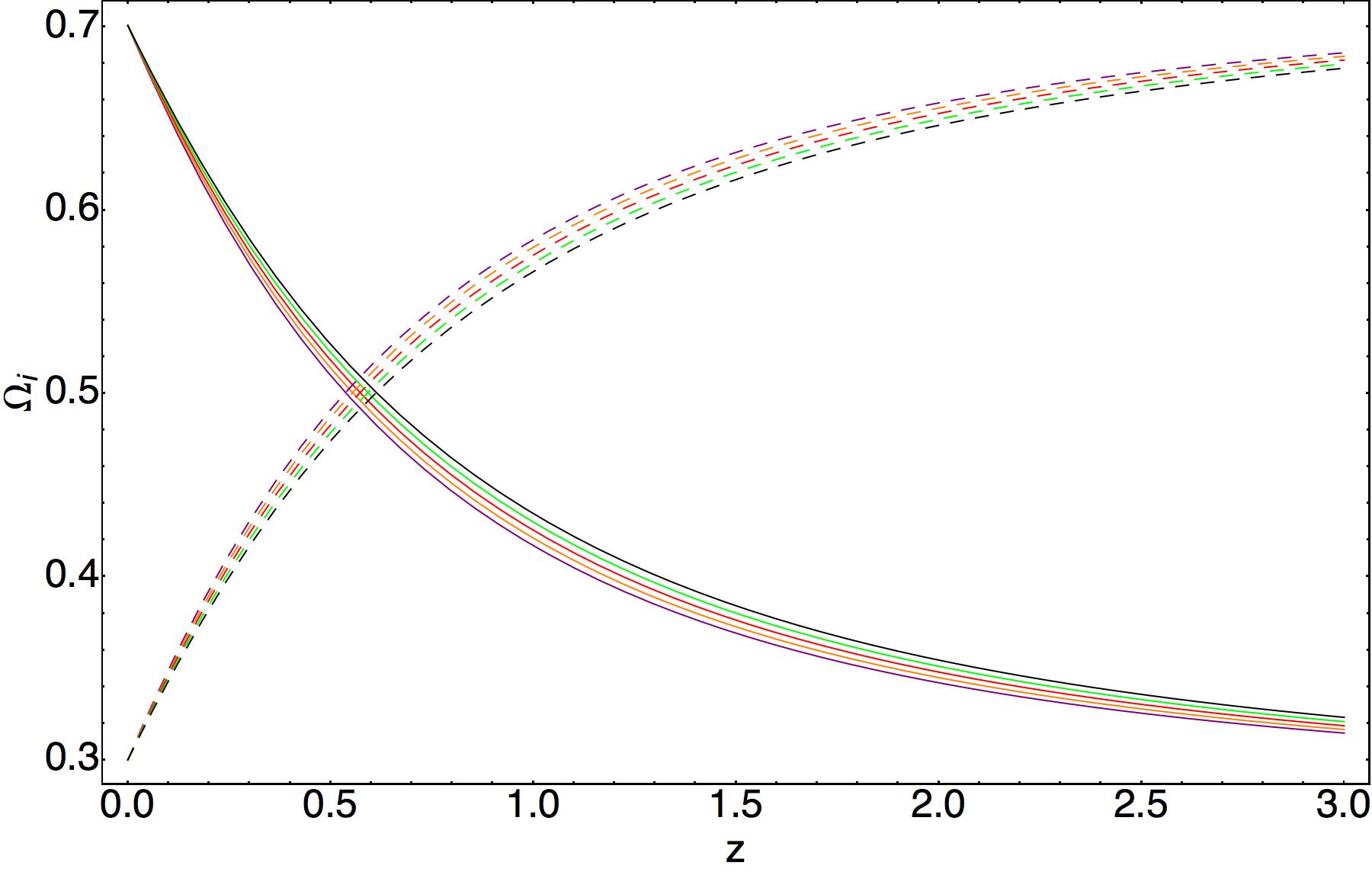}  \\
 \end{array}$
 \end{center}
\caption{Graphical behavior of the EoS parameter $\omega_{tot}$ of the effective fluid and $\Omega_{i}$ against redshift $z$. $m=0$ does correspond to usual ghost dark energy. Behavior of $\Omega_{de}$ is represented by a blue curve on $\Omega_{i} - z$ plane, while orange line does represent behavior of $\Omega_{dm}$ (for $m=-0.2$).  Considered behavior does correspond to interacting varying ghost dark energy Eq.~(\ref{eq:VGDE}), when the interaction is given via Eq.~(\ref{eq:INT1}).}
 \label{fig:Fig1_1}
\end{figure}

First interacting varying ghost dark energy model considered in this paper, is the model with the interaction of the following form
\begin{equation}\label{eq:INT1}
Q  = 3 b H (\rho_{de}+ \rho_{dm}),
\end{equation}
where $b$ it is a positive constant, $H$ it is the Hubble parameter. It is obvious, that consideration of an interaction will modify results discussed in previous section. Particularly, in case of considered interaction, Eq~(\ref{eq:INT1}), the EoS parameter and the deceleration parameter will be modified appropriately as prescribed bellow
\begin{equation}\label{eq:omegadeINT1}
\omega_{de} = \frac{-\Omega _{\text{de}} \left(6 b (m-1)+\beta +(6 m-3) \Omega _{\text{de}}-2 \beta  m-6 m+3\right)+2 b (\beta  m-3)+\beta -2 \beta  m}{\left(\Omega _{\text{de}}-1\right) \Omega _{\text{de}} \left(\beta +3 \Omega _{\text{de}}-6\right)},
\end{equation}
\begin{equation}\label{eq:qINT1}
q = -\frac{3 (2 m-1) (3 b-\beta +3)}{\beta +3 \Omega _{\text{de}}-6}+\frac{3 b m}{\Omega _{\text{de}}-1}-3 m+2.
\end{equation}
It is easy to see that for $\Omega_{de} \to 0$, $\omega_{de} \to \infty$, but $q$ is finite and it is equal
\begin{equation}
q = -\frac{3 (2 m-1) (3 b-\beta +3)}{\beta -3} - 3 b m -3 m + 2,
\end{equation} 
on the other hand, when $\Omega_{de} \to 1$, then $\omega_{de}$ and $q$ tend to infinity, therefore for this model $\Omega_{de} = 1$ case should be excluded from our future consideration. Starting from the first law of thermodynamics, after some mathematics, we will have the differential equation describing the dynamics of entropy of the varying ghost dark energy presented bellow
\begin{equation}
T\frac{dS_{\text{de}}}{dH} = \frac{4 \pi}{3H^{2}} \left ( \frac{18 (2 m-1) (3 b-\beta +3)}{\beta +3 \Omega _{\text{de}}-6}-\frac{18 b m}{\Omega _{\text{de}}-1}+ A + A_{1} \right ),
\end{equation}
where 
\begin{equation}
A = \frac{(\beta -3)^2 m (2 m-1)}{(m-1)^2 \left(-3 (m-1) \Omega _{\text{de}}+\beta  m-3\right)},
\end{equation}
and 
\begin{equation}
A_{1} = \frac{3 (2-3 m) \Omega _{\text{de}}}{m-1}+\frac{(2 m-1) (-\beta +3 m (3 m-5)+9)}{(m-1)^2}. 
\end{equation}
Fig.~(\ref{fig:Fig1}) and Fig.~(\ref{fig:Fig1_1}) represent behavior of the deceleration parameter $q$, the EoS parameter of varying ghost dark energy, the EoS parameter of the effective fluid and behavior of $\Omega_{de}$ and $\Omega_{dm}$ corresponding to different values of the parameter $b$ for fixed values of $\beta$ and $m$. In this way, we will try to understand how considered interaction will act on the behavior of mentioned parameters. Discussed values for $\beta$ and $m$ give the best fit for the theoretical results with the distance modulus $\mu$. From Fig.~(\ref{fig:Fig1}) and Fig.~(\ref{fig:Fig1_1}), we see that an increase of the  value of the parameter of interaction term will only decrease $q$, $\omega_{de}$ and $\omega_{tot}$ as well as will decrease present day values of these parameters. Moreover, we see that with considered increase of $b$, the transition redshift giving us the accelerated expanding recent universe will increase. Left plot of Fig.~(\ref{fig:Fig1_1}) demonstrates graphical behavior of $\Omega_{de}$ and $\Omega_{dm}$ for different values the parameters of the model identical to these which have been used for $q$, $\omega_{de}$ and $\omega_{tot}$. It can be seen easily, that an increase of $b$ brings to an increase $\Omega_{de}$~(solid lines) and to a decrease of $\Omega_{dm}$~(dashed lines). To complete this subsection, in Table.~(\ref{tab:Table1}) we have summarized present day values of the statefinder pair $(r,s)$ and $(\omega^{\prime}_{de},\omega_{de})$ for different values of $b$. Recall, that $(\omega^{\prime}_{de},\omega_{de})$ it is another tool developed in order to distinguish dark energy models.

\begin{table}
  \centering
    \begin{tabular}{ | l | l | l | l | p{1cm} |}
    \hline
 $b$ & $(r,s)$ & $(\omega_{de}^{\prime}, \omega_{de})$\\
    \hline
 $0.00$ & $(5.250, -1.636)$ & $(-1.896, -0.824)$ \\
    \hline
 $0.01$ & $(4.991, -1.499)$ & $(-1.784, -0.844)$ \\
    \hline
 $0.02$ & $(4.739, -1.372)$ & $(-1.675, -0.865)$ \\
    \hline
 $0.03$ & $(4.495, -1.253)$ & $(-1.569, -0.885)$ \\
    \hline
 $0.04$ & $(4.261, -1.143)$ & $(-1.467, -0.905)$ \\
    \hline
    \end{tabular}
\caption{Present day values of $(r,s)$ and $(\omega_{de}^{\prime}, \omega_{de})$ for various values of parameter $b$, when $\alpha = 0.75$, $\beta=0.84$, $m=-0.2$., $H_{0}=0.7$, $\Omega_{de}\approx 0.7$ and $\Omega_{dm} \approx 0.3$. The interaction is given via Eq.~(\ref{eq:INT1}).}
  \label{tab:Table1}
\end{table}

\subsection{Interacting model $2$}\label{ssec:INT2}
\begin{figure}[h!]
 \begin{center}$
 \begin{array}{cccc}
 \includegraphics[width=80 mm]{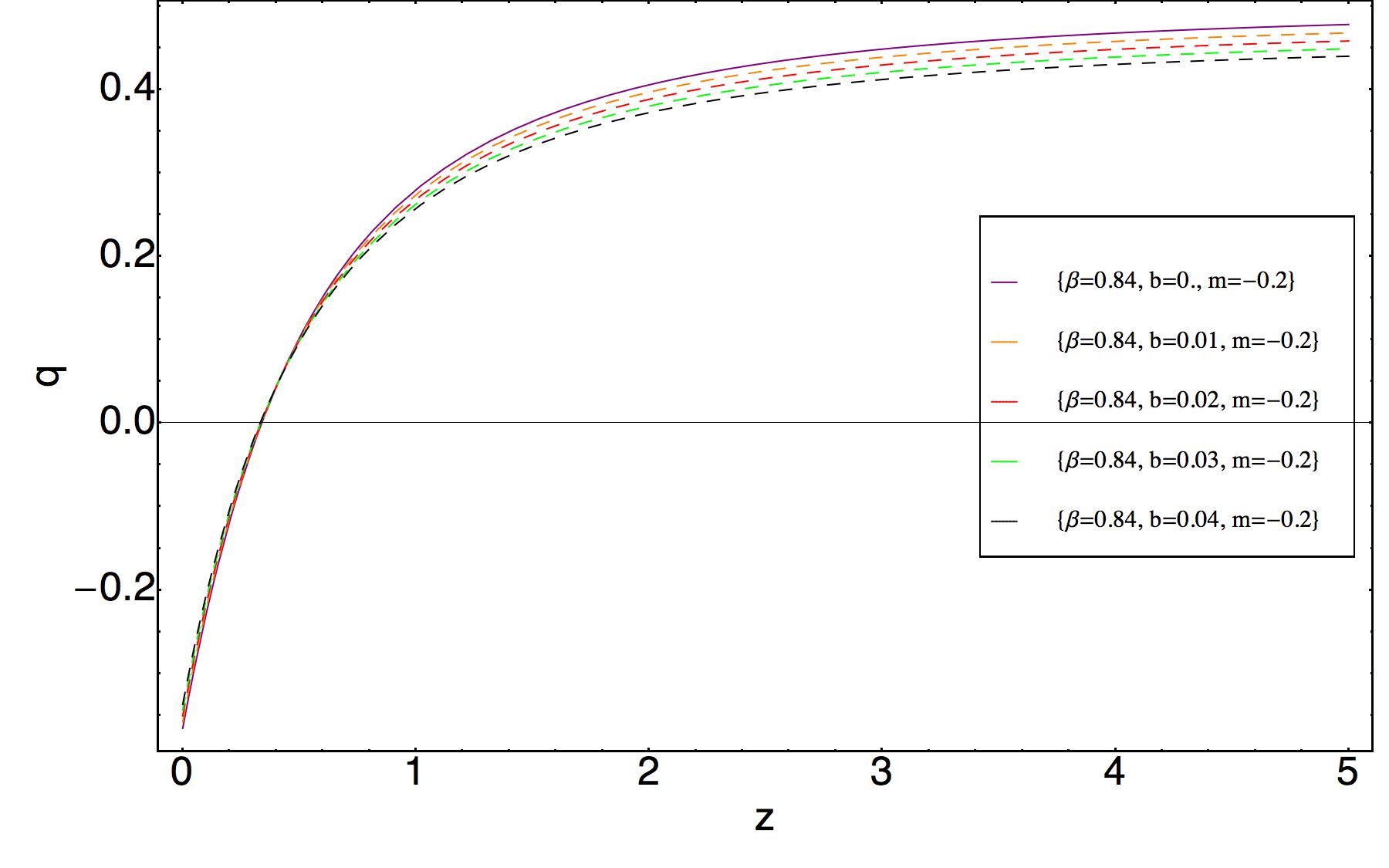}  &
\includegraphics[width=80 mm]{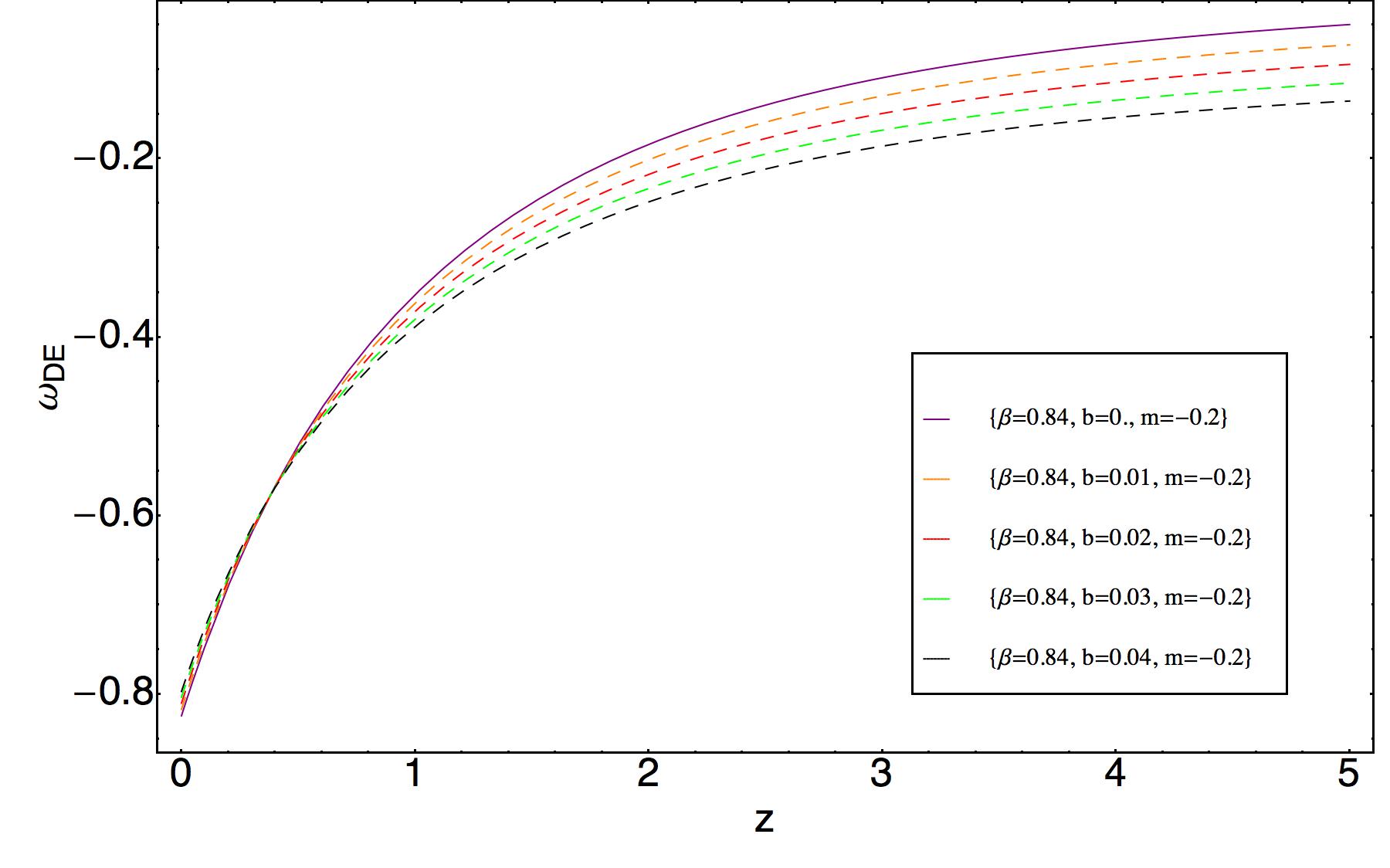}  \\
 \end{array}$
 \end{center}
\caption{Graphical behavior of the deceleration parameter $q$ and $\omega_{de}$ of interacting varying ghost dark energy Eq.~(\ref{eq:VGDE}) against redshift $z$. $m=0$ does correspond to usual ghost dark energy. The interaction is given via Eq.~(\ref{eq:INT2}).}
 \label{fig:Fig2}
\end{figure}

\begin{figure}[h!]
 \begin{center}$
 \begin{array}{cccc}
 \includegraphics[width=80 mm]{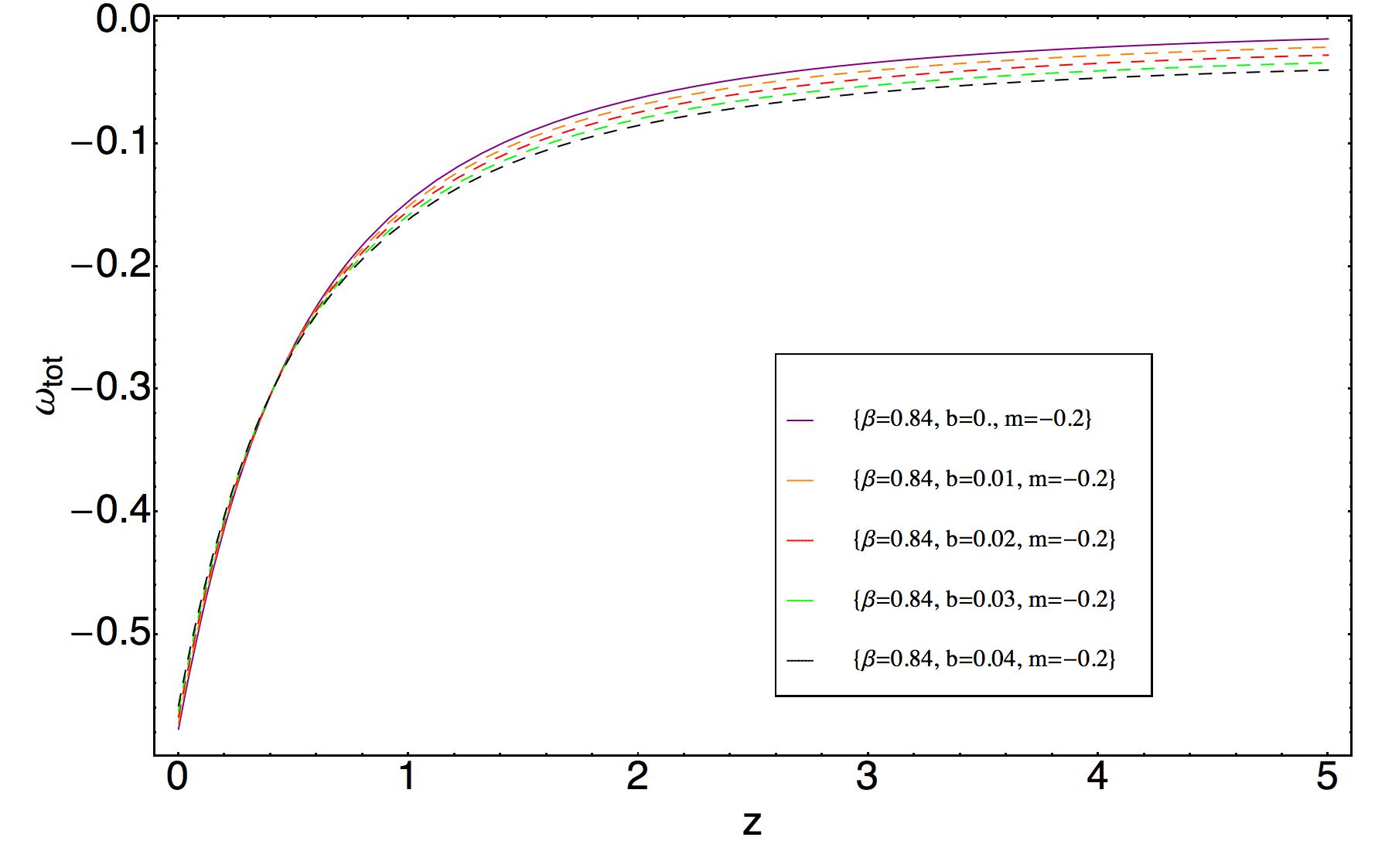}  &
\includegraphics[width=80 mm]{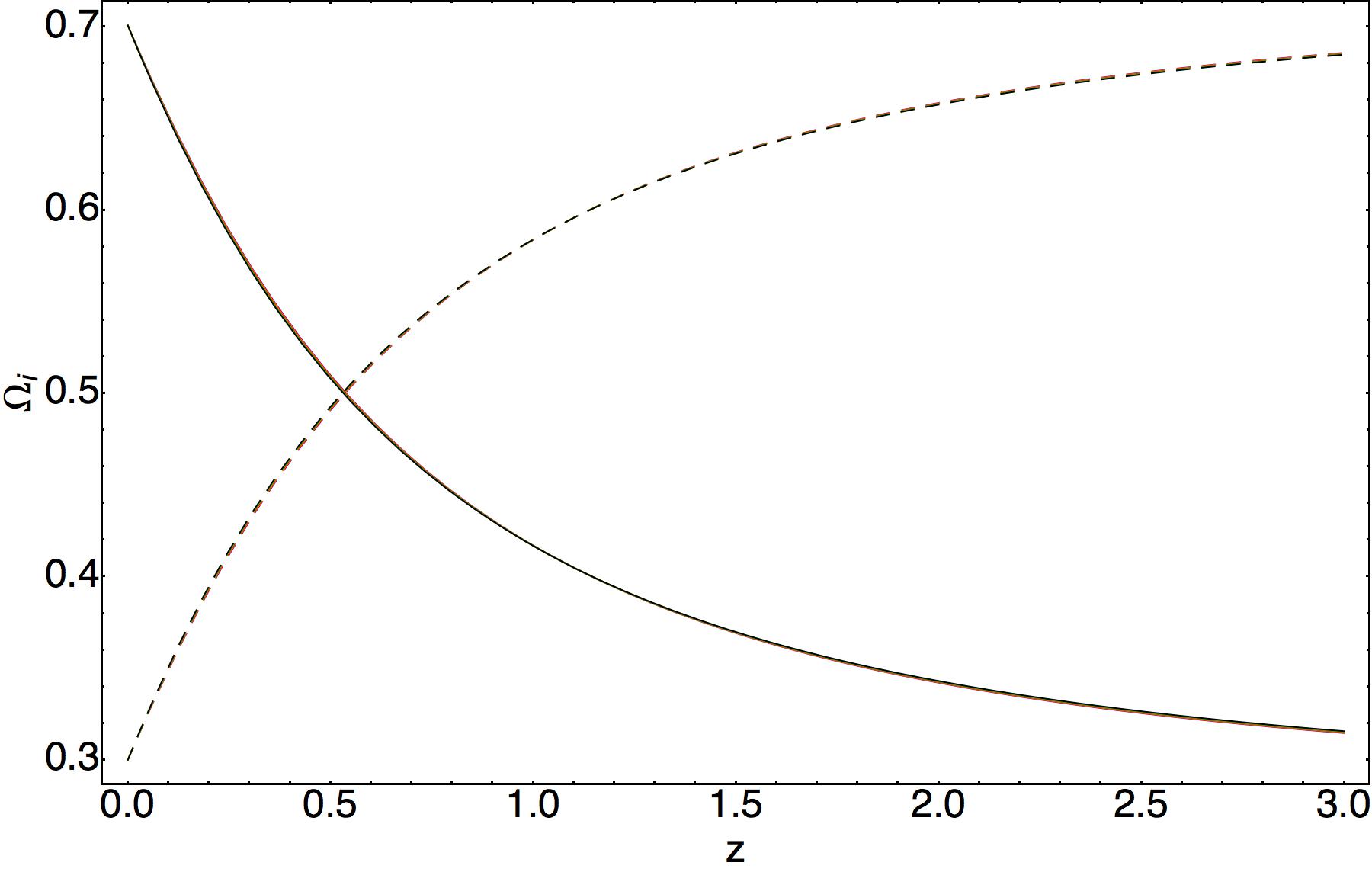}  \\
 \end{array}$
 \end{center}
\caption{Graphical behavior of the EoS parameter $\omega_{tot}$ of the effective fluid and $\Omega_{i}$ against redshift $z$. $m=0$ does correspond to usual ghost dark energy. Behavior of $\Omega_{de}$ is represented by a blue curve on $\Omega_{i} - z$ plane, while orange line does represent behavior of $\Omega_{dm}$ (for $m=-0.2$).  Considered behavior does correspond to interacting varying ghost dark energy Eq.~(\ref{eq:VGDE}), when the interaction is given via Eq.~(\ref{eq:INT2}).}
 \label{fig:Fig2_1}
\end{figure}

The sign changeable interaction which we will consider to describe another cosmological model involving suggested varying ghost dark energy, it is one of the first models of such kind of interaction considered in Literature. Generally, there is not any theoretical and observational fact against to this idea. Moreover, from a point of view of phenomenology it is quit interesting to study cosmological models with sign changeable interactions, because sometimes completely different possibilities interesting from point of view of cosmology can be observed. The model of the sign changeable interaction considered in this subsection involves the deceleration parameter $q$ in order to provide sign changeability and has the following form   

\begin{equation}\label{eq:INT2}
Q  = 3 b H q (\rho_{de}+ \rho_{dm})
\end{equation}
In Fig.~(\ref{fig:Fig2}) and Fig.~(\ref{fig:Fig2_1}) the reader can find redshift dependent behavior of the deceleration parameter $q$, the EoS parameter of varying ghost dark energy, the EoS parameter of the effective fluid and behavior of $\Omega_{de}$ and $\Omega_{dm}$ corresponding to different values of the parameter $b$ for fixed values of $\beta$ and $m$. We can see that considered sign changeable interaction, Eq.~(\ref{eq:INT2}), does not affect on transition redshift, moreover, present day values of the parameters for different values of $b$ are close enough, and practically, interacting models can not be distinguished from the non interacting model. Therefore, statefinder analysis, $(\omega_{de}^{\prime}, \omega_{de})$ and $Om$ analysis could be used together to have clarification of the situation to which we will come in next section. Estimation of the present day values of $(r,s)$ and $(\omega_{de}^{\prime}, \omega_{de})$~(Table~\ref{tab:Table2}) for this model give us almost comparable results with non interacting model~(for small values of $b$), therefore we are left only with $Om$ analysis to obtain more information. Considered model, it is a model of the universe where 
\begin{equation}\label{eq:omegadeINT2}
\omega_{de} = \frac{-\Omega _{\text{de}} \left(3 b (m-1)+\beta +(6 m-3) \Omega _{\text{de}}-2 \beta  m-6 m+3\right)+b (\beta  m-3)+\beta -2 \beta  m}{\Omega _{\text{de}} \left(\Omega _{\text{de}} \left(9 b (m-1)+\beta +3 \Omega _{\text{de}}-9\right)+b (9-3 \beta  m)-\beta +6\right)},
\end{equation}
\begin{equation}\label{eq:qINT2}
q =-\frac{\left(\Omega _{\text{de}}-1\right) \left(\beta +(9 m-6) \Omega _{\text{de}}-3 \beta  m+3\right)}{\Omega _{\text{de}} \left(9 b (m-1)+\beta +3 \Omega _{\text{de}}-9\right)+b (9-3 \beta  m)-\beta +6}.
\end{equation}
and the dynamics of entropy of the varying ghost dark energy, Eq.~(\ref{eq:VGDE}), according to the first law of thermodynamics has the following form
\begin{equation}\label{eq:dSdHINT2}
T \frac{dS_{\text{de}}}{dH} = \frac{4 \pi  \left(\Omega _{\text{de}} \left(\beta +(9 m-6) \Omega _{\text{de}}-3 (\beta +2) m+6\right)+\beta  (2 m-1)\right)}{H^2 \left(-3 (m-1) \Omega _{\text{de}}+\beta  m-3\right)}-\frac{12 \pi  \omega _{\text{de}}  \Omega _{\text{de}}}{H^2},
\end{equation}
where $\omega_{de}$ is given by Eq.~(\ref{eq:omegadeINT2}). From Eq.~(\ref{eq:qINT2}), we see that when $\Omega_{de} = 1$, then $q = 0$, therefore to have the large scale universe with the accelerated expansion with $q \in [-1,0)$, $\omega_{de} \in [-1,0)$ and for simplicity $0 \leq b <1$ with $0 \leq \beta <1$, we need to have
\begin{equation}
\frac{1}{3}<\Omega _{\text{de}}<1,
\end{equation}
and 
\begin{equation}
\frac{\left(\Omega _{\text{de}}-1\right) \left(\Omega _{\text{de}} \left(-9 b+\beta +3 \Omega _{\text{de}}-3\right)+3 b-\beta \right)}{\left((3 b-2) \Omega _{\text{de}}-b+2\right) \left(\beta -3 \Omega _{\text{de}}\right)}\leq m<\frac{\beta -6 \Omega _{\text{de}}+3}{3 \beta -9 \Omega _{\text{de}}}.
\end{equation}
For the universe with $\Omega_{de} = 0$ i.e. matter dominated universe we have
\begin{equation}
q =\frac{\left(\beta -3 \beta  m+3\right)}{b (9-3 \beta  m)-\beta +6}.
\end{equation}

\begin{table}
  \centering
    \begin{tabular}{ | l | l | l | l | p{1cm} |}
    \hline
 $b$ & $(r,s)$ & $(\omega_{de}^{\prime}, \omega_{de})$\\
    \hline
 $0.00$ & $(5.250, -1.636)$ & $(-1.896, -0.824)$ \\
    \hline
 $0.01$ & $(5.217, -1.638)$ & $(-1.826, -0.817)$ \\
    \hline
 $0.02$ & $(5.184, -1.639)$ & $(-1.756, -0.810)$ \\
    \hline
 $0.03$ & $(5.149, -1.638)$ & $(-1.686, -0.804)$ \\
    \hline
 $0.04$ & $(5.114, -1.637)$ & $(-1.616, -0.797)$ \\
    \hline
    \end{tabular}
\caption{Present day values of $(r,s)$ and $(\omega_{de}^{\prime}, \omega_{de})$ for various values of parameter $b$, when $\alpha = 0.75$, $\beta=0.84$, $m=-0.2$., $H_{0}=0.7$, $\Omega_{de}\approx 0.7$ and $\Omega_{dm} \approx 0.3$. The interaction is given via Eq.~(\ref{eq:INT2}).}
  \label{tab:Table2}
\end{table}

\subsection{Interacting model $3$}\label{ssec:INT3}
\begin{figure}[h!]
 \begin{center}$
 \begin{array}{cccc}
 \includegraphics[width=80 mm]{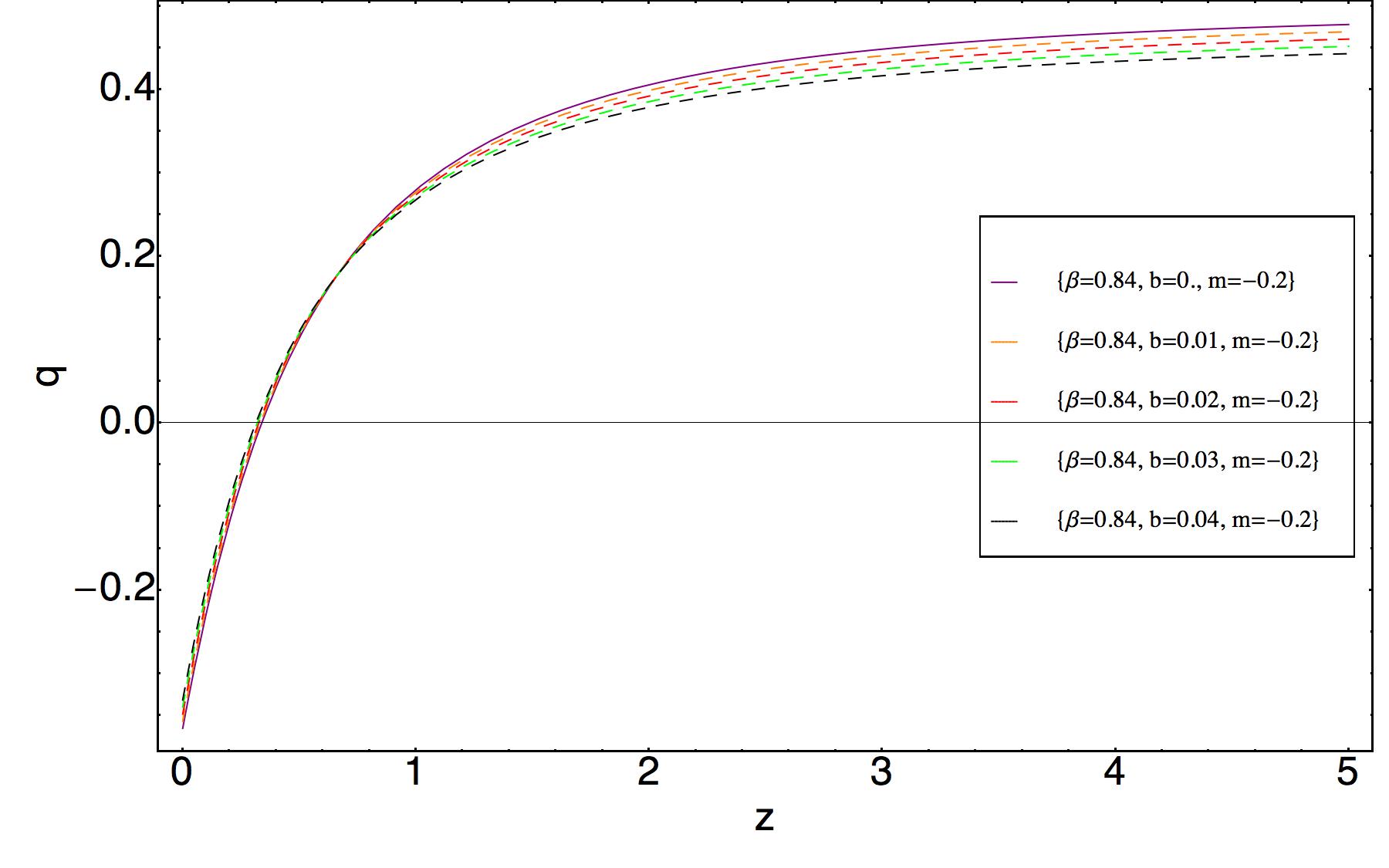}  &
  \includegraphics[width=80 mm]{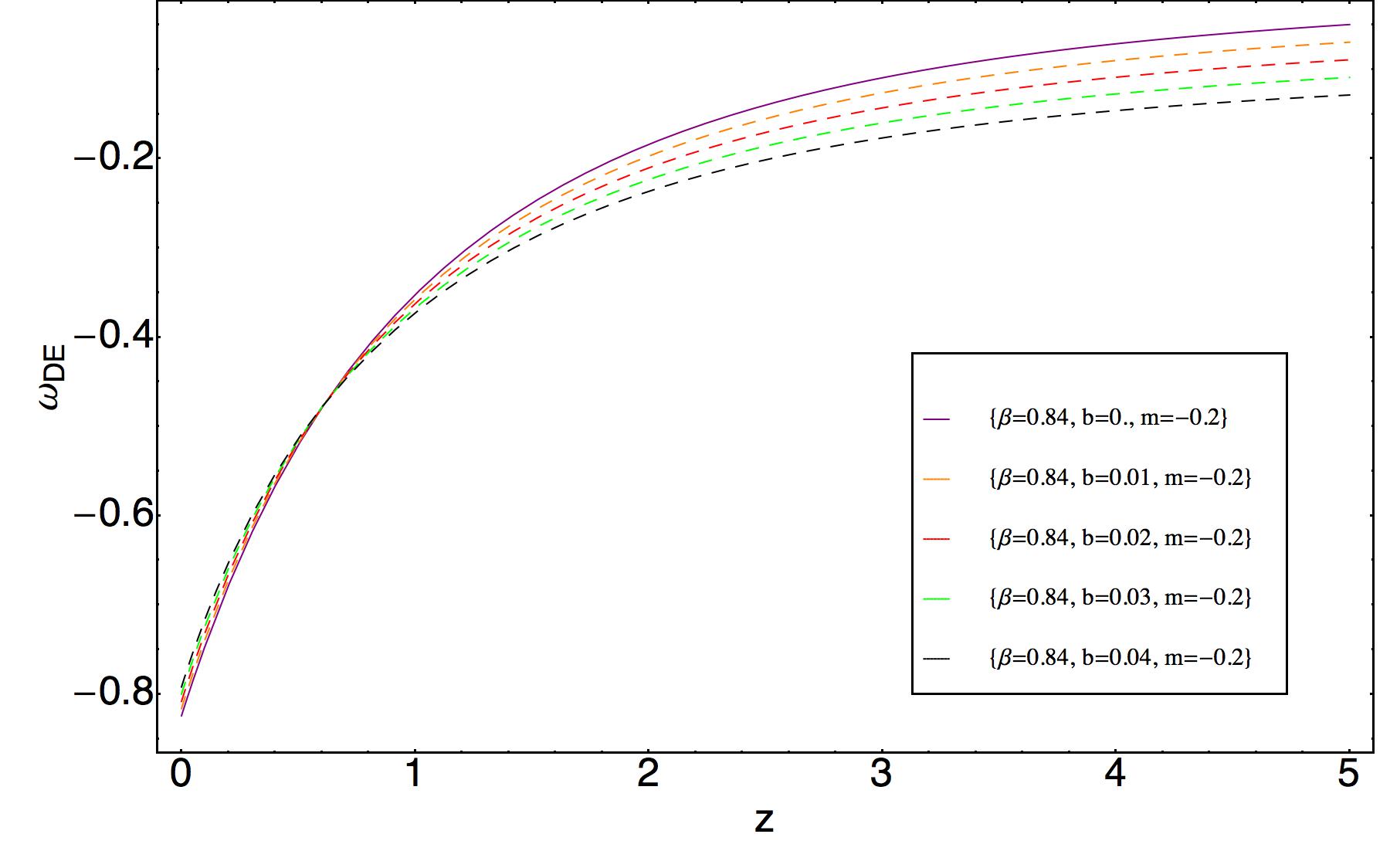}  \\
\includegraphics[width=80 mm]{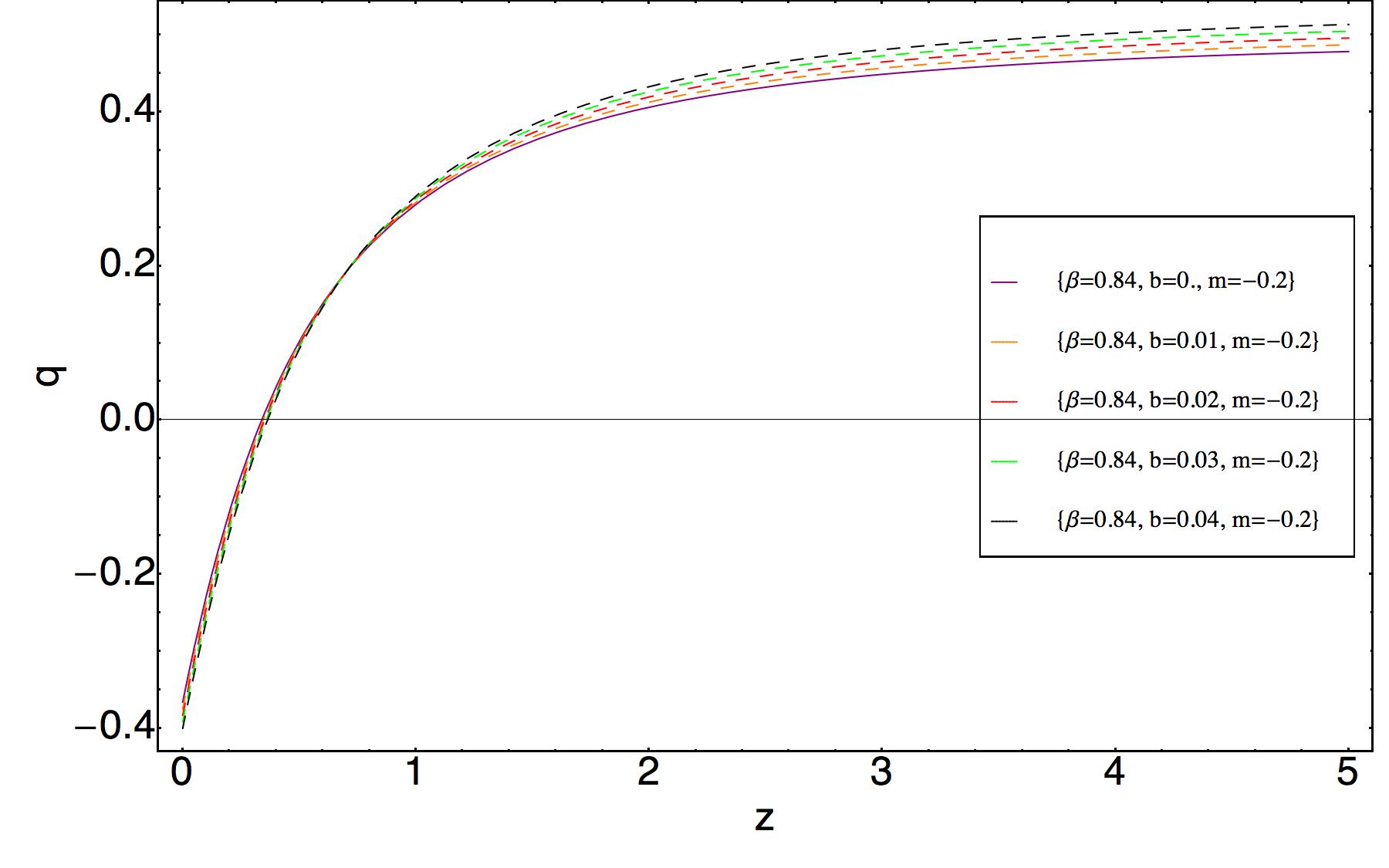}  &
\includegraphics[width=80 mm]{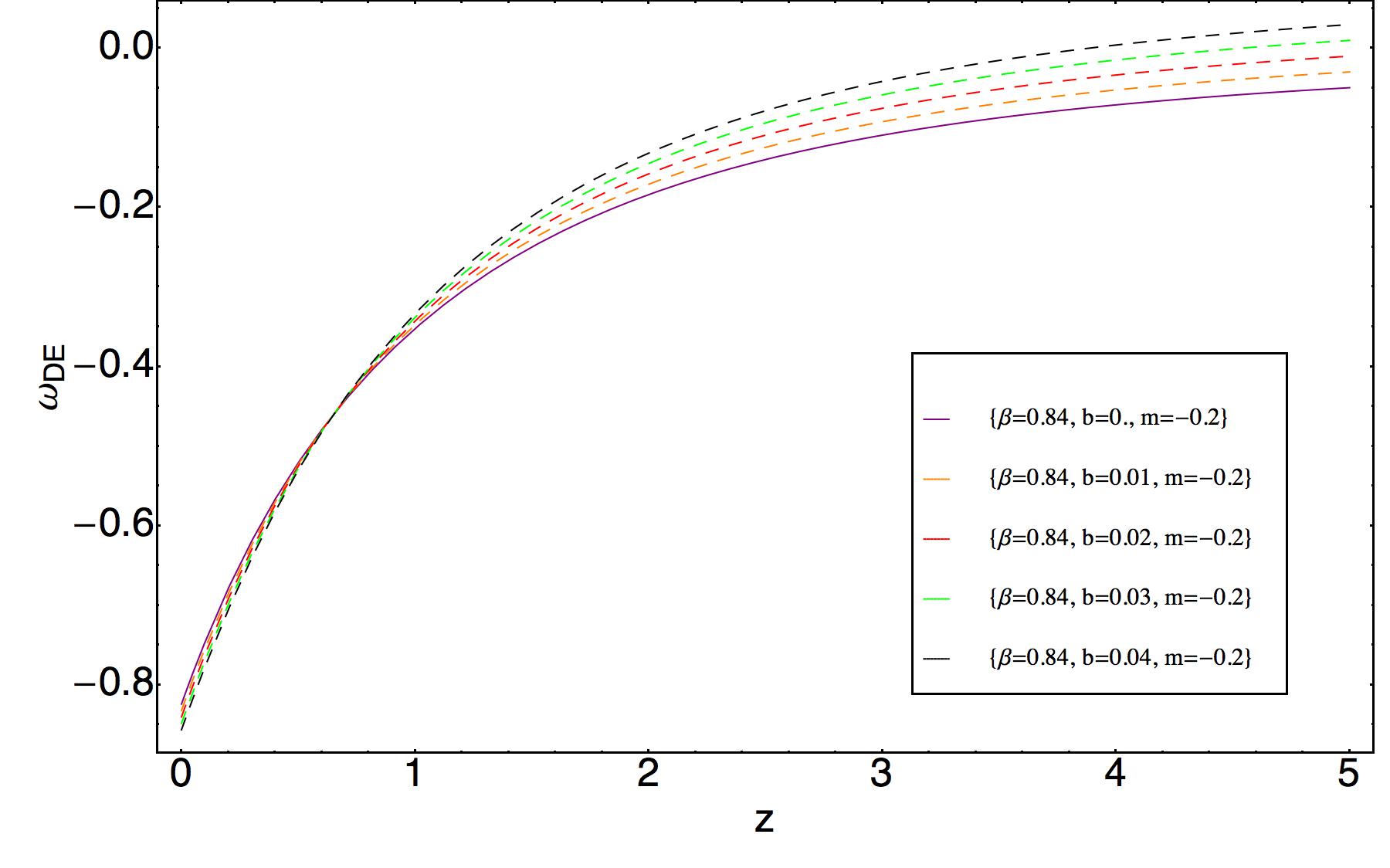}
 \end{array}$
 \end{center}
\caption{Graphical behavior of the deceleration parameter $q$ and $\omega_{de}$ of interacting varying ghost dark energy Eq.~(\ref{eq:VGDE}) against redshift $z$. $m=0$ does correspond to usual ghost dark energy. Interaction is given via Eq.~(\ref{eq:INT3}).}
 \label{fig:Fig3}
\end{figure}

\begin{figure}[h!]
 \begin{center}$
 \begin{array}{cccc}
 \includegraphics[width=80 mm]{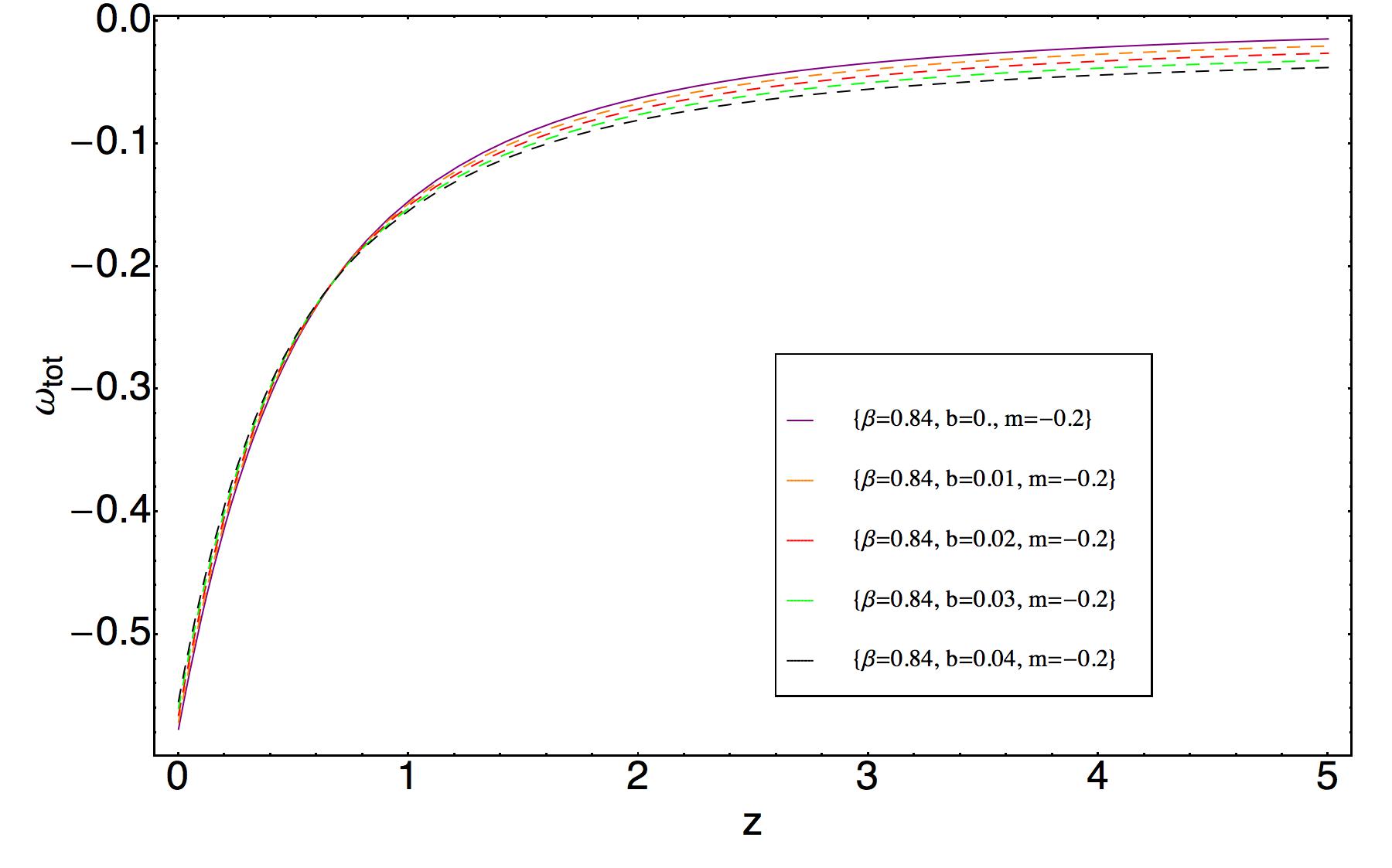}  &
 \includegraphics[width=80 mm]{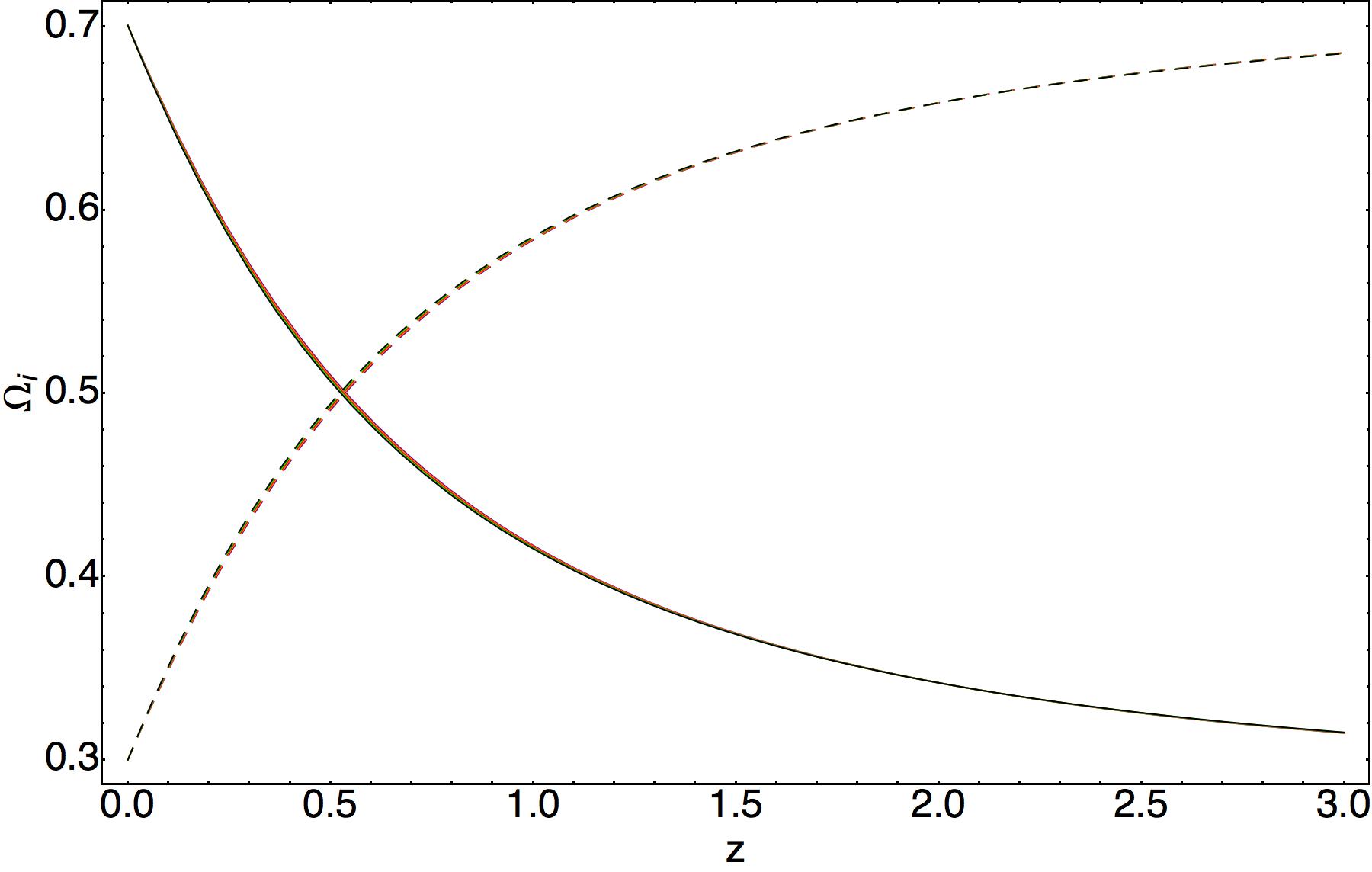} \\
\includegraphics[width=80 mm]{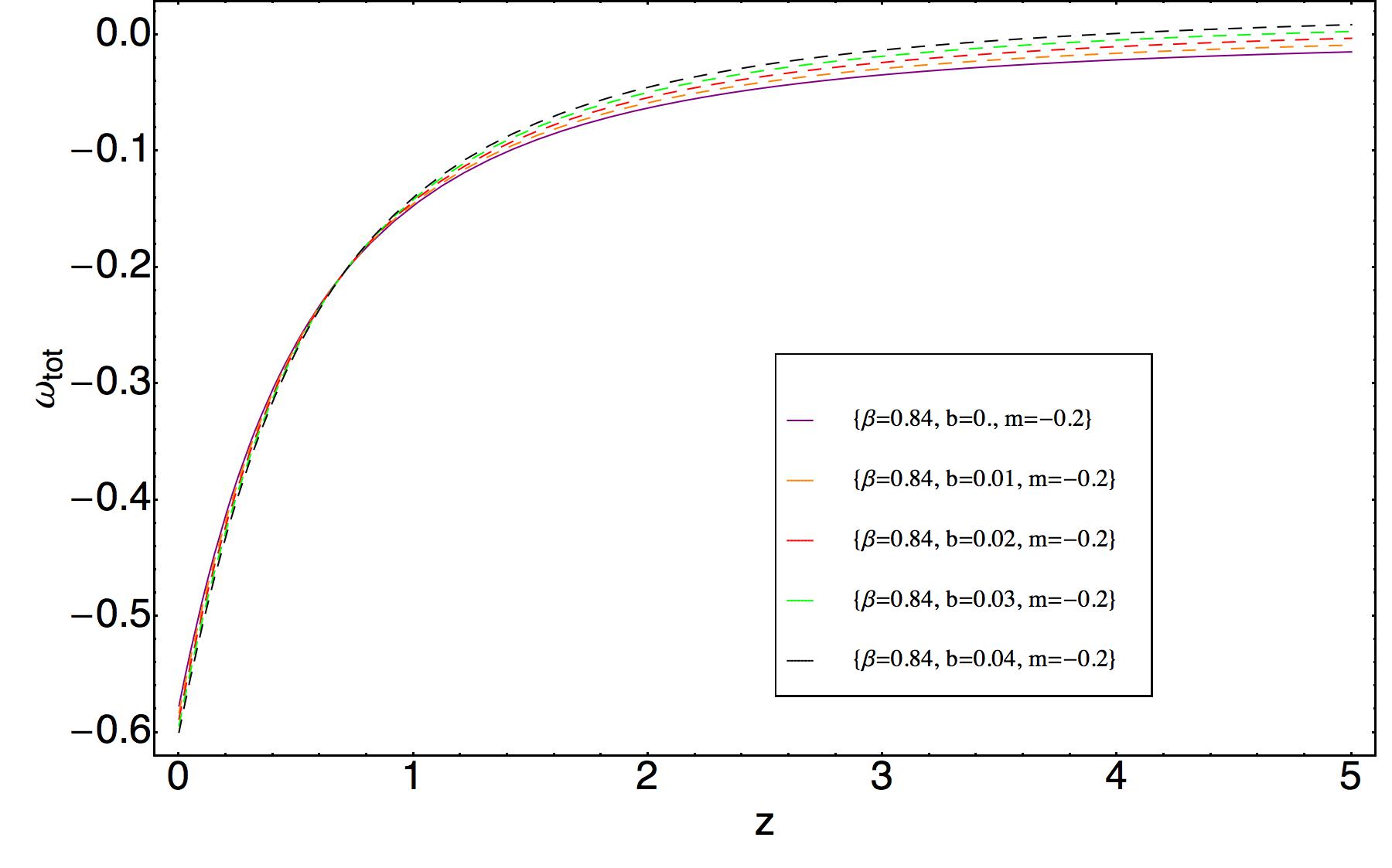}  &
\includegraphics[width=80 mm]{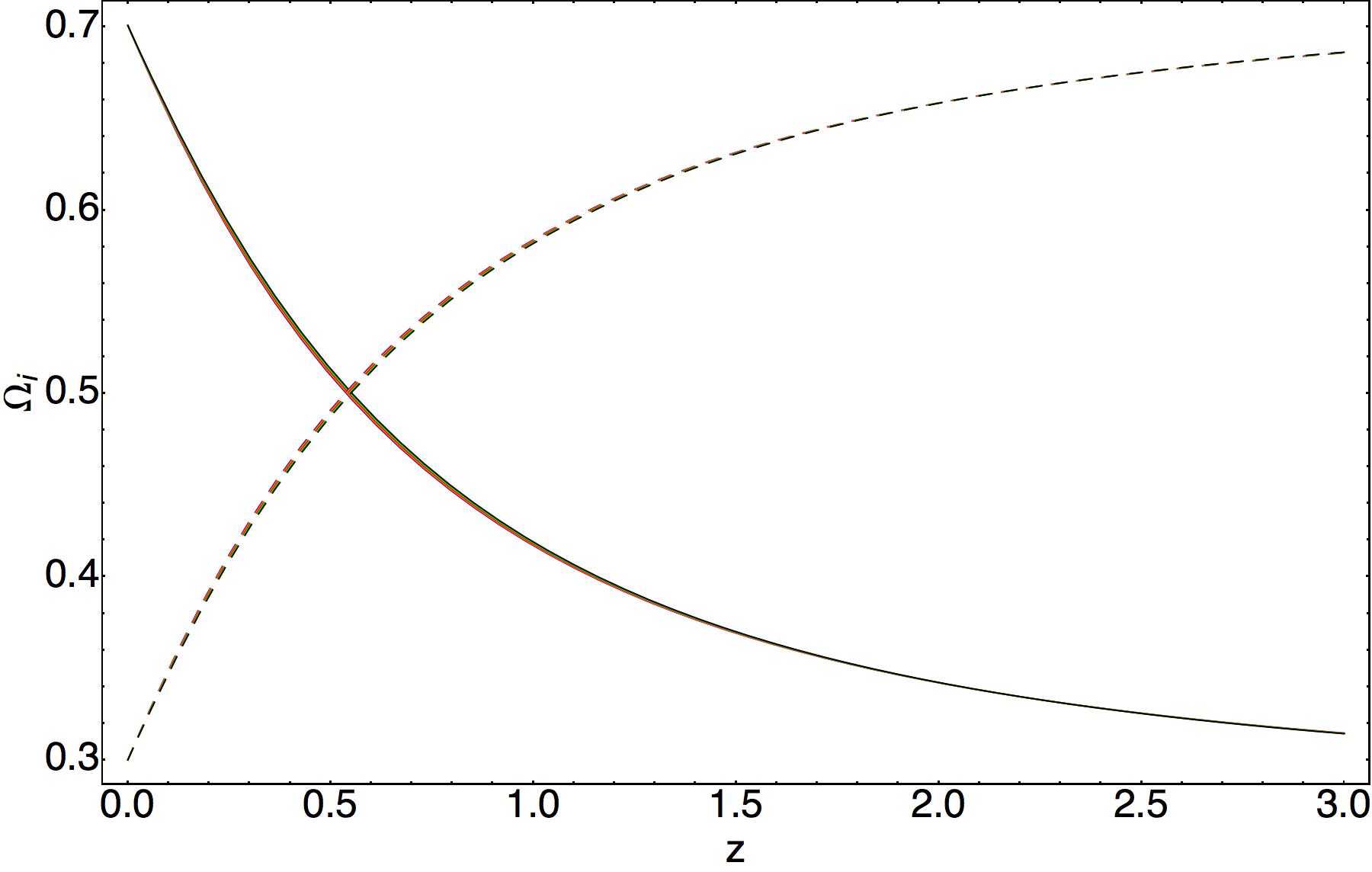} 
 \end{array}$
 \end{center}
\caption{Graphical behavior of the EoS parameter $\omega_{tot}$ of the effective fluid and $\Omega_{i}$ against redshift $z$. $m=0$ does correspond to usual ghost dark energy. Behavior of $\Omega_{de}$ is represented by a blue curve on $\Omega_{i} - z$ plane, while orange line does represent behavior of $\Omega_{dm}$ (for $m=-0.2$).  Considered behavior does correspond to interacting varying ghost dark energy Eq.~(\ref{eq:VGDE}), when the interaction is given via Eq.~(\ref{eq:INT3}).}
 \label{fig:Fig3_1}
\end{figure}

Discussed sign changeable interaction Eq.~(\ref{eq:INT2}), it is one of the options. Actually, the difference between the energy densities of the dark energy and dark matter can be used to achieve to a desirable result, like in the case presented bellow   
\begin{equation}\label{eq:INT3}
Q  = 3 b H (\rho_{dm} - \rho_{de}).
\end{equation}
According to this model of sign changeable interaction, when $\rho_{de} > \rho_{dm}$ dark matter will transfer to dark energy. In opposite case dark energy will transfer to dark matter. Consideration of the interaction Eq.~(\ref{eq:INT3}) gives us a universe, where the EoS parameter has the following form
\begin{equation}\label{eq:omegadeINT3}
\omega_{de} = \frac{A_{3} +2 b (\beta  m-3)+\beta -2 \beta  m}{\left(\Omega _{\text{de}}-1\right) \Omega _{\text{de}} \left(\beta +3 \Omega _{\text{de}}-6\right)},
\end{equation}
giving the following form to the deceleration parameter $q$
\begin{equation}\label{eq:qINT3}
q = -\frac{3 (2 m-1) (b (2 \beta -9)-\beta +3)}{\beta +3 \Omega _{\text{de}}-6}-\frac{3 b m}{\Omega _{\text{de}}-1}+6 b m-6 b-3 m+2,
\end{equation}
where $A_{3} = \Omega _{\text{de}} \left(3 (4 b (m-1)-2 m+1) \Omega _{\text{de}}-2 b (2 \beta  m+3 m-9)+(\beta +3) (2 m-1)\right)$. To have the large scale universe with the accelerated expansion with $q \in [-1,0)$, $\omega_{de} \in [-1,0)$ and for simplicity $0 \leq b <1$ with $0 \leq \beta <1$, we need to have
\begin{equation}
\frac{1}{3}<\Omega _{\text{de}}<1
\end{equation}
\begin{equation}
\frac{\left(\Omega _{\text{de}}-1\right) \left(\Omega _{\text{de}} \left(-12 b+\beta +3 \Omega _{\text{de}}-3\right)+6 b-\beta \right)}{2 \left((2 b-1) \Omega _{\text{de}}-b+1\right) \left(\beta -3 \Omega _{\text{de}}\right)}\leq m<\frac{\left(\Omega _{\text{de}}-1\right) \left(6 (3 b-1) \Omega _{\text{de}}-9 b+\beta +3\right)}{3 \left((1-2 b) \Omega _{\text{de}}+b-1\right) \left(\beta -3 \Omega _{\text{de}}\right)}.
\end{equation}
Redshift dependent behavior of the deceleration parameter $q$, the EoS parameter $\omega_{de}$, the EoS of the effective fluid $\omega_{tot}$ with the behavior of $\Omega_{de}$ and $\Omega_{dm}$ can be found in Fig.~(\ref{fig:Fig3}) and Fig.~(\ref{fig:Fig3_1}), respectively. Considered model it is a cosmological model where a transition form a decelerated expanding universe to the accelerated expending recent large scale universe is possible due to correct behavior of considered cosmological parameters. It is obvious, that we could consider another model of interaction when we will use $\rho_{de} - \rho_{dm}$. The difference between these two models it is the sign of parameter $b$, therefore if in discussed equations of this section we will consider $|b|$ instead of $b$ then we can have physics describing of two models depending on the sign of $b$. We used this and during graphical study of mentioned cosmological parameters we will compare results related to these two models of interaction combining them in  Fig.~(\ref{fig:Fig3}) and Fig.~(\ref{fig:Fig3_1}). From the top panel of Fig.~(\ref{fig:Fig3}) we see that for high redshifts interaction Eq.~(\ref{eq:INT3}) will decrease the deceleration parameter $q$ (top-left plot), while consideration of the interaction $Q=3Hb(\rho_{de} - \rho_{dm})$ will increase the same parameter (top-right plot). On the other hand, transition redshift for the first case will decrease with an increasing of the interaction parameter $b$, while for the second case it will decrease. Moreover, for low redshifts interaction will play an opposite role compared to its role for hig redshifts i.e. for the first case the deceleration parameter will increase (bottom-left plot), while will decrease for the second case with interaction $Q=3Hb(\rho_{de} - \rho_{dm})$. Bottom panel of the same plot and the top panel of Fig.~(\ref{fig:Fig3_1}) demonstrate appropriate behavior of $\omega_{de}$ and $\omega_{tot}$. On the other hand, the bottom panel of Fig.~(\ref{fig:Fig3_1}) indicates that impact of considered sign changeable interactions on $\Omega_{de}$ and $\Omega_{dm}$ can be neglected with high accuracy i.e. practically interaction does not leave any information into the dynamics of $\Omega_{de}$ and $\Omega_{dm}$. Combining two types of interaction considered in this subsection into one, we found that the dynamics of entropy of the varying ghost dark energy has the following form 
\begin{equation}\label{eq:dSdHINT3}
T\frac{dS_{\text{de}}}{dH} = \frac{4 \pi  \left(\Omega _{\text{de}} \left(\beta +(9 m-6) \Omega _{\text{de}}-3 (\beta +2) m+6\right)+\beta  (2 m-1)\right)}{H^2 \left(-3 (m-1) \Omega _{\text{de}}+\beta  m-3\right)}-\frac{12 \pi  \omega_{\text{de}}  \Omega _{\text{de}}}{H^2},
\end{equation}
where $\omega_{\text{de}}$ is given by Eq.~(\ref{eq:omegadeINT3}) and where interaction parameter $b$ has been changed to $|b|$ respectively. Present day values of $(r,s)$ and $(\omega_{de}^{\prime}, \omega_{de})$ parameters for both cases are presented in Table~\ref{tab:Table3}  and Table~\ref{tab:Table4}, respectively. 

\begin{table}
  \centering
    \begin{tabular}{ | l | l | l | l | p{1cm} |}
    \hline
 $b$ & $(r,s)$ & $(\omega_{de}^{\prime}, \omega_{de})$\\
    \hline
 $0.00$ & $(5.250, -1.636)$ & $(-1.896, -0.824)$ \\
    \hline
 $0.01$ & $(5.267, -1.659)$ & $(-1.902, -0.816)$ \\
    \hline
 $0.02$ & $(5.283, -1.682)$ & $(-1.908, -0.808)$ \\
    \hline
 $0.03$ & $(5.298, -1.704)$ & $(-1.913, -0.800)$ \\
    \hline
 $0.04$ & $(5.312, -1.727)$ & $(-1.918, -0.792)$ \\
    \hline
    \end{tabular}
\caption{Present day values of $(r,s)$ and $(\omega_{de}^{\prime}, \omega_{de})$ for various values of parameter $b$, when $\alpha = 0.75$, $\beta=0.84$, $m=-0.2$., $H_{0}=0.7$, $\Omega_{de}\approx 0.7$ and $\Omega_{dm} \approx 0.3$.The interaction is given by Eq.~(\ref{eq:INT3}).}
  \label{tab:Table3}
\end{table}

\begin{table}
  \centering
    \begin{tabular}{ | l | l | l | l | p{1cm} |}
    \hline
 $b$ & $(r,s)$ & $(\omega_{de}^{\prime}, \omega_{de})$\\
    \hline
 $0.00$ & $(5.250, -1.636)$ & $(-1.896, -0.824)$ \\
    \hline
 $0.01$ & $(5.232, -1.613)$ & $(-1.889, -0.832)$ \\
    \hline
 $0.02$ & $(5.213, -1.590)$ & $(-1.882, -0.841)$ \\
    \hline
 $0.03$ & $(5.193, -1.568)$ & $(-1.875, -0.848)$ \\
    \hline
 $0.04$ & $(5.172, -1.545)$ & $(-1.866, -0.856)$ \\
    \hline
    \end{tabular}
\caption{Present day values of $(r,s)$ and $(\omega_{de}^{\prime}, \omega_{de})$ for various values of parameter $b$, when $\alpha = 0.75$, $\beta=0.84$, $m=-0.2$., $H_{0}=0.7$, $\Omega_{de}\approx 0.7$ and $\Omega_{dm} \approx 0.3$. The interaction is given by $Q=3Hb(\rho_{de} - \rho_{dm})$.}
  \label{tab:Table4}
\end{table}

\section{$Om$ and statefinder hierarchy of the models}\label{sec:PSA}
\begin{figure}[h!]
 \begin{center}$
 \begin{array}{cccc}
 \includegraphics[width=80 mm]{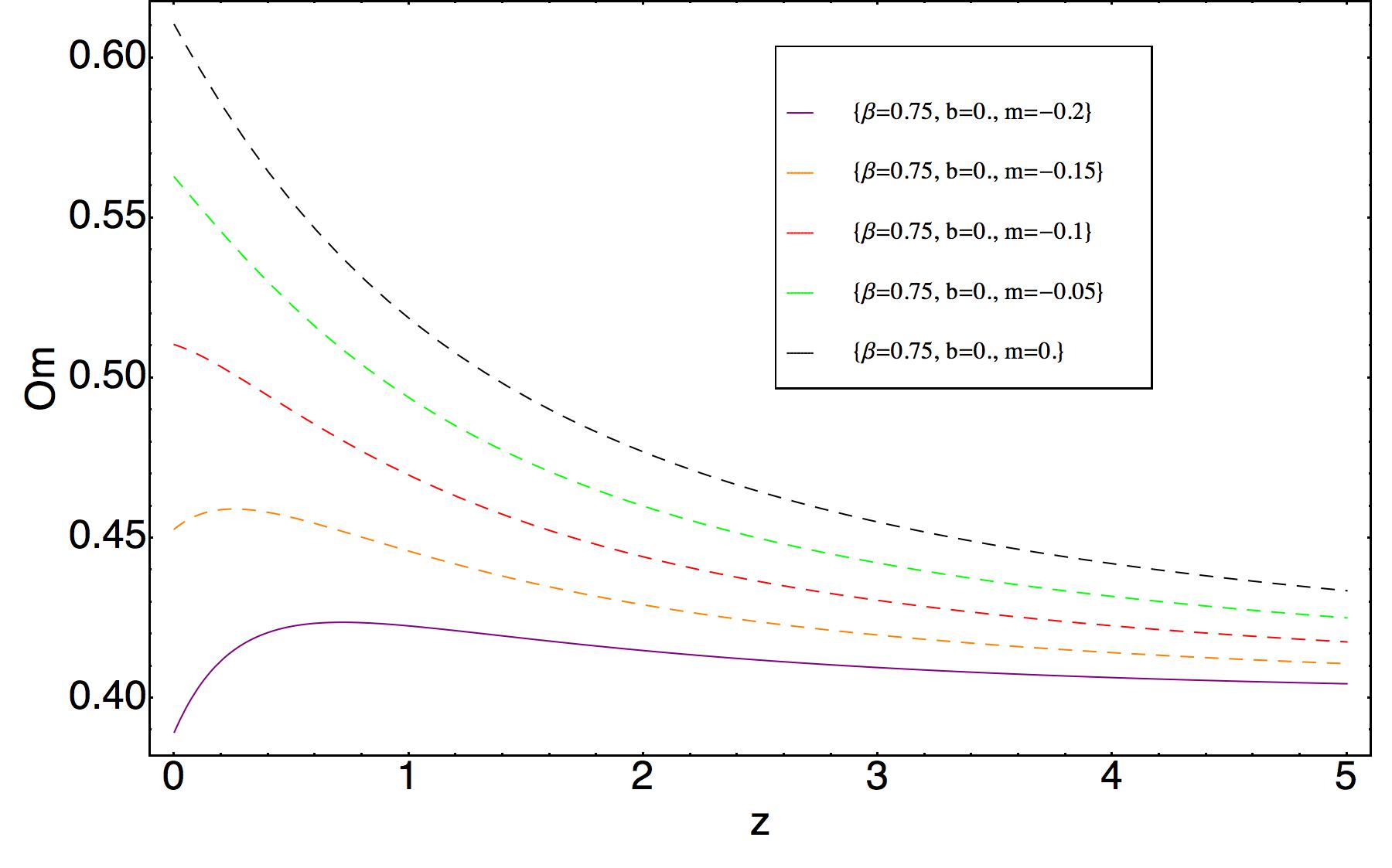}  &
\includegraphics[width=80 mm]{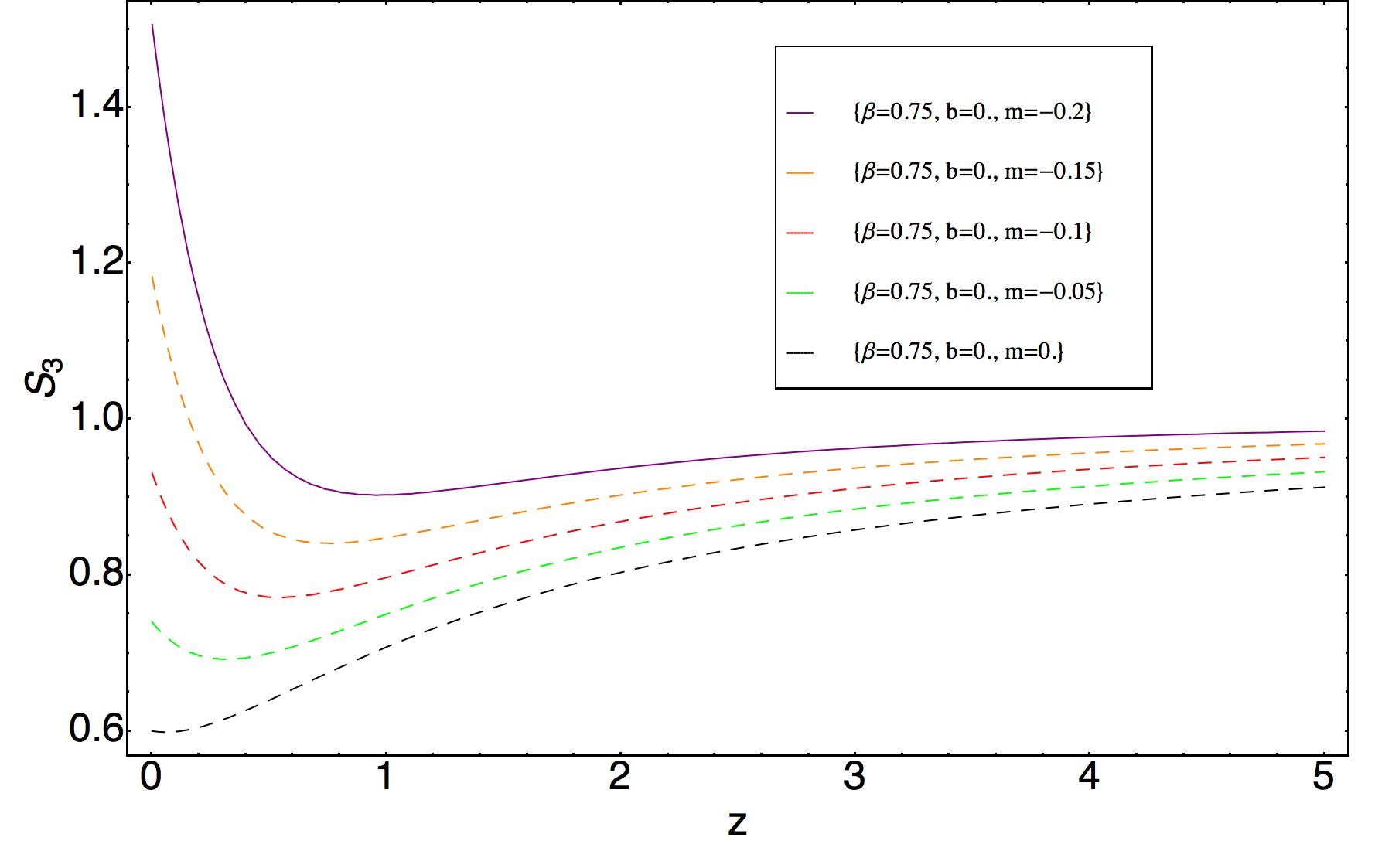}  \\
 \end{array}$
 \end{center}
\caption{Graphical behavior of $Om$ and $S_{3}$ parameters against redshift $z$. $m=0$ does correspond to usual ghost dark energy. Analysis presents cosmological model with non interacting varying ghost dark energy Eq.~(\ref{eq:VGDE}).}
 \label{fig:Fig4}
\end{figure}

\begin{figure}[h!]
 \begin{center}$
 \begin{array}{cccc}
 \includegraphics[width=80 mm]{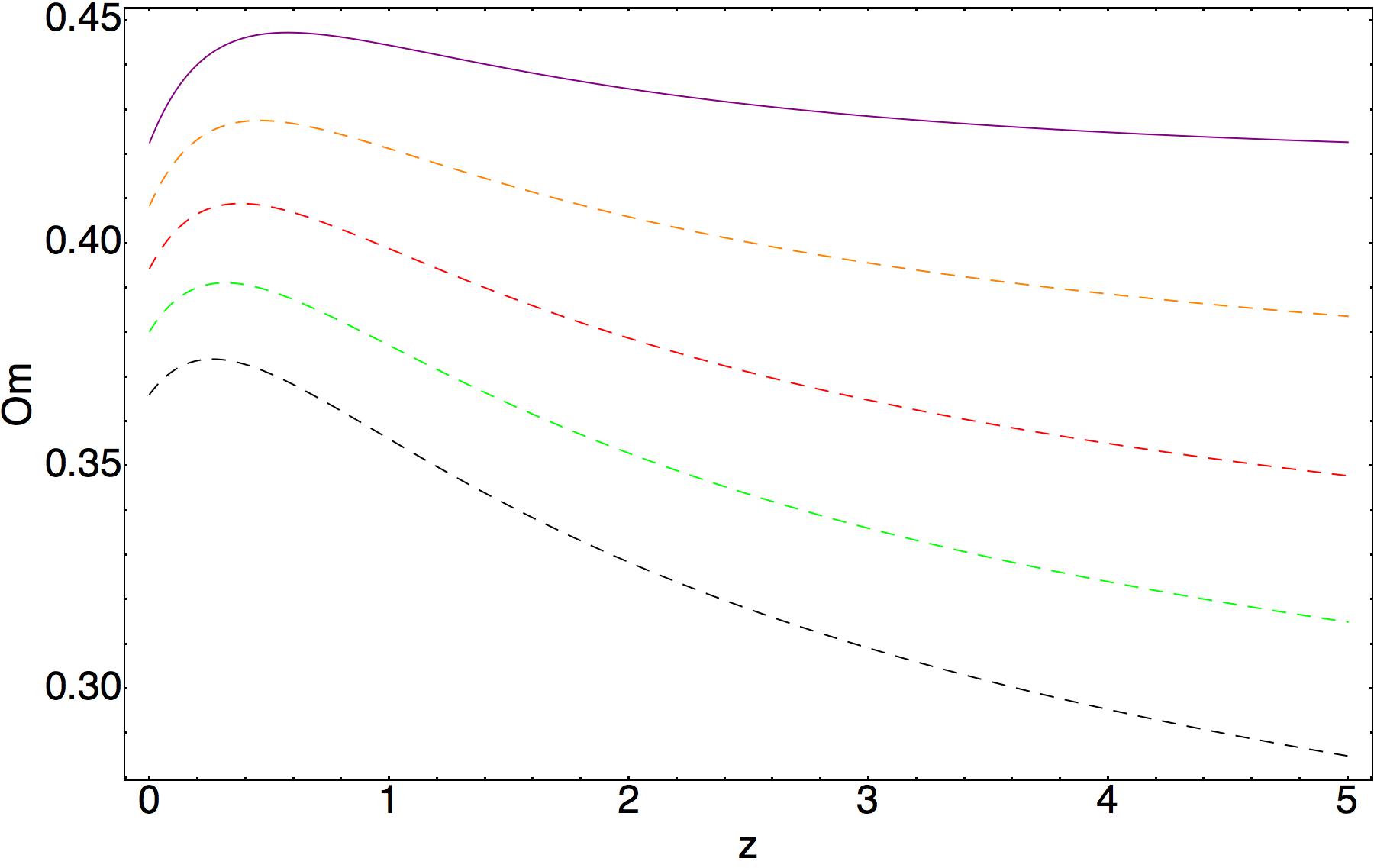}  &
\includegraphics[width=80 mm]{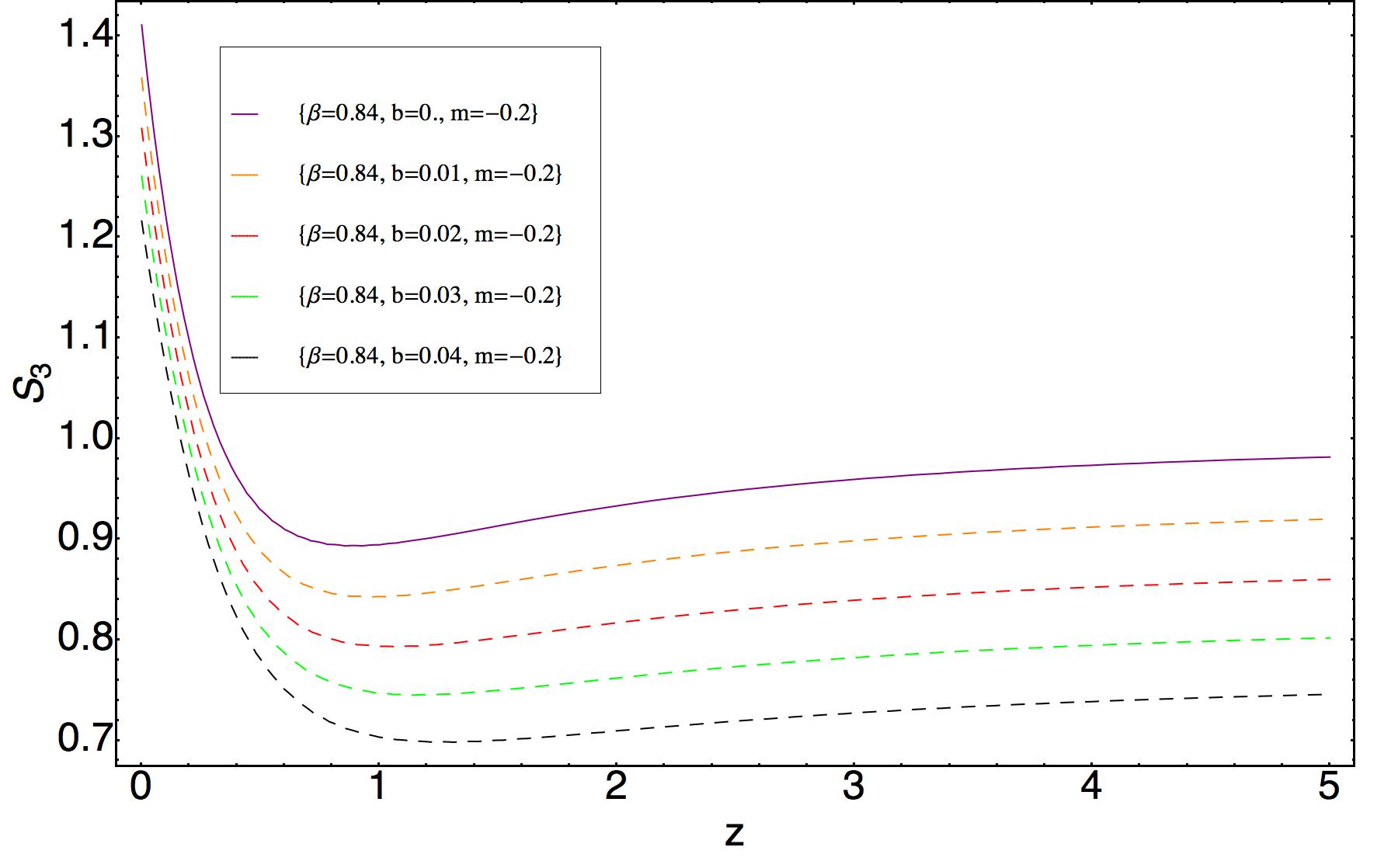}  \\
 \end{array}$
 \end{center}
\caption{Graphical behavior of $Om$ and $S_{3}$ parameters against redshift $z$. $m=0$ does correspond to usual ghost dark energy. Analysis presents cosmological model with interacting varying ghost dark energy Eq.~(\ref{eq:VGDE}). The interaction is given via Eq.~(\ref{eq:INT1}). Behavior of $Om$ parameter is according to the same values of the parameters of the model as for $S_{3}$ parameter.}
 \label{fig:Fig5}
\end{figure}

\begin{figure}[h!]
 \begin{center}$
 \begin{array}{cccc}
 \includegraphics[width=80 mm]{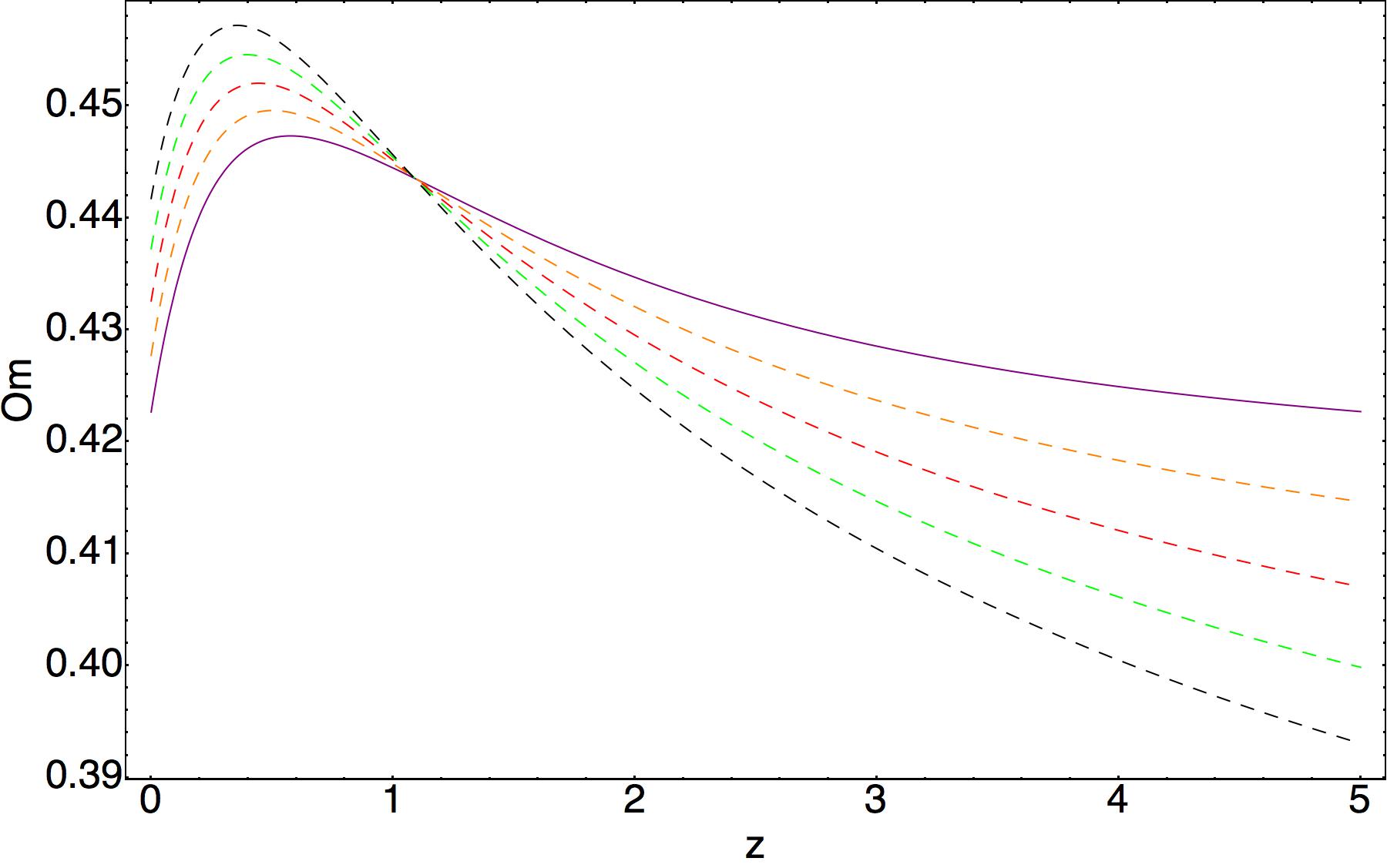}  &
\includegraphics[width=80 mm]{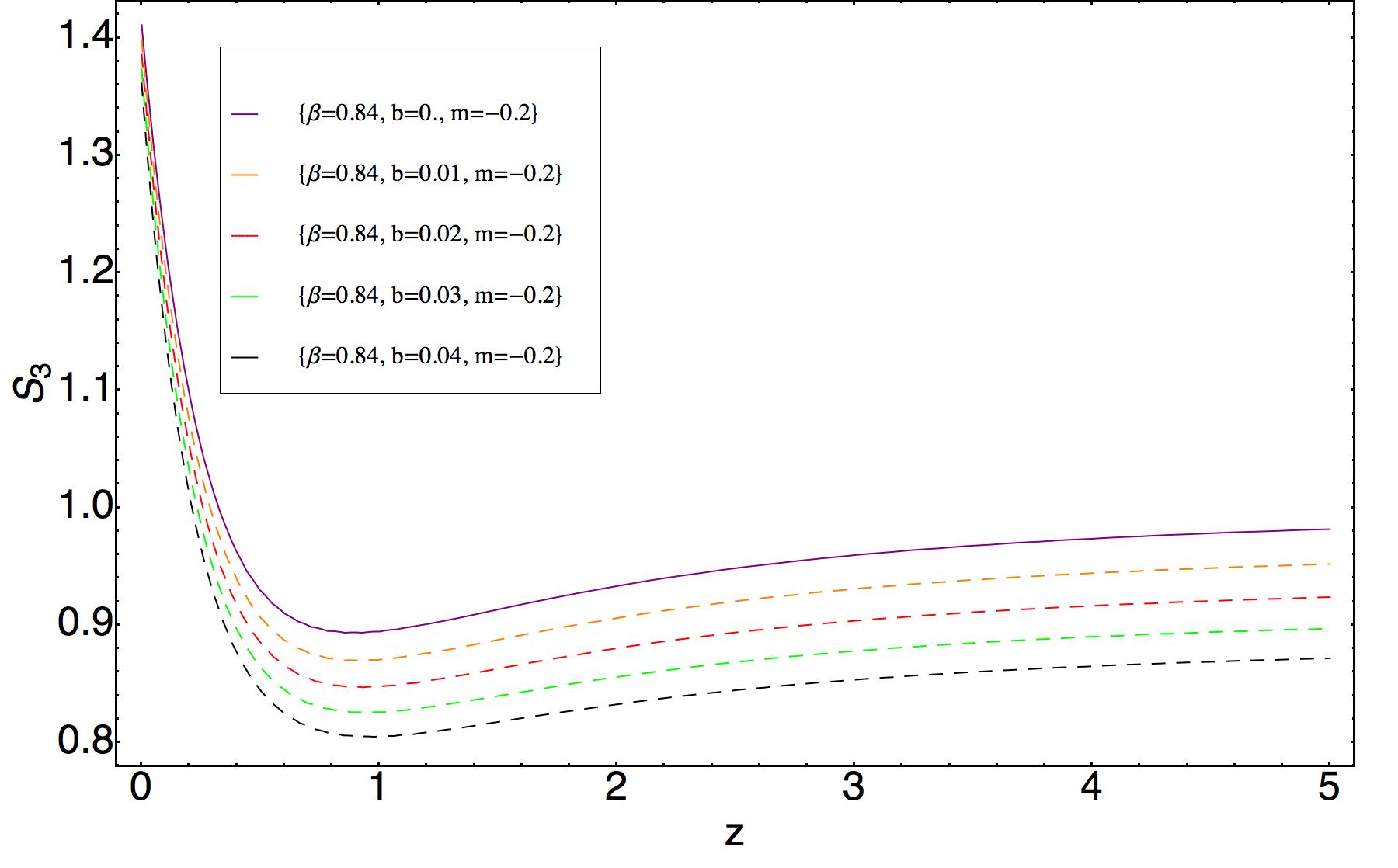}  \\
 \end{array}$
 \end{center}
\caption{Graphical behavior of $Om$ and $S_{3}$ parameters against redshift $z$. $m=0$ does correspond to usual ghost dark energy. Analysis presents cosmological model with interacting varying ghost dark energy Eq.~(\ref{eq:VGDE}). The interaction is given via Eq.~(\ref{eq:INT2}). Behavior of $Om$ parameter is according to the same values of the parameters of the model as for $S_{3}$ parameter.}
 \label{fig:Fig6}
\end{figure}

\begin{figure}[h!]
 \begin{center}$
 \begin{array}{cccc}
 \includegraphics[width=80 mm]{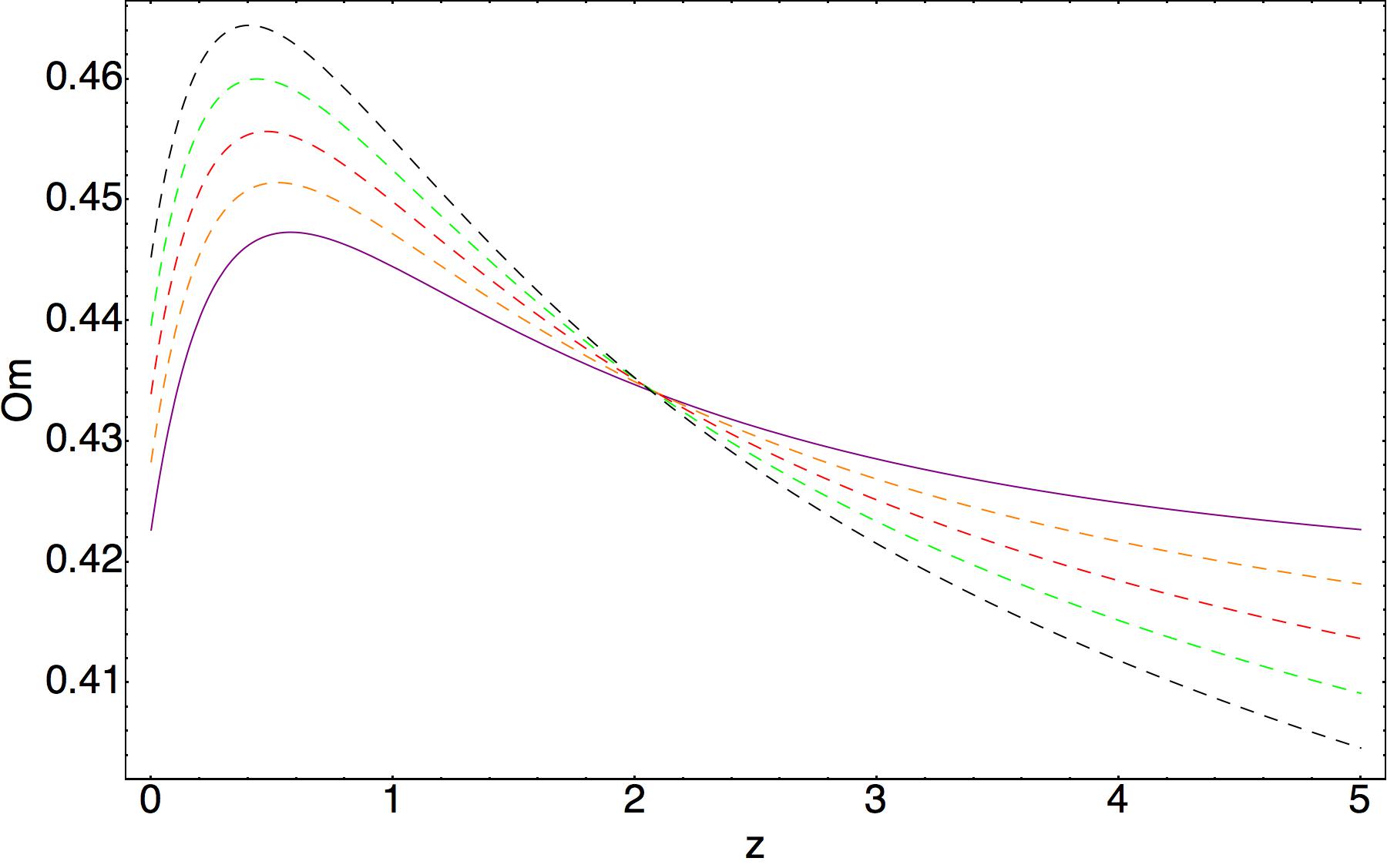}  &
\includegraphics[width=80 mm]{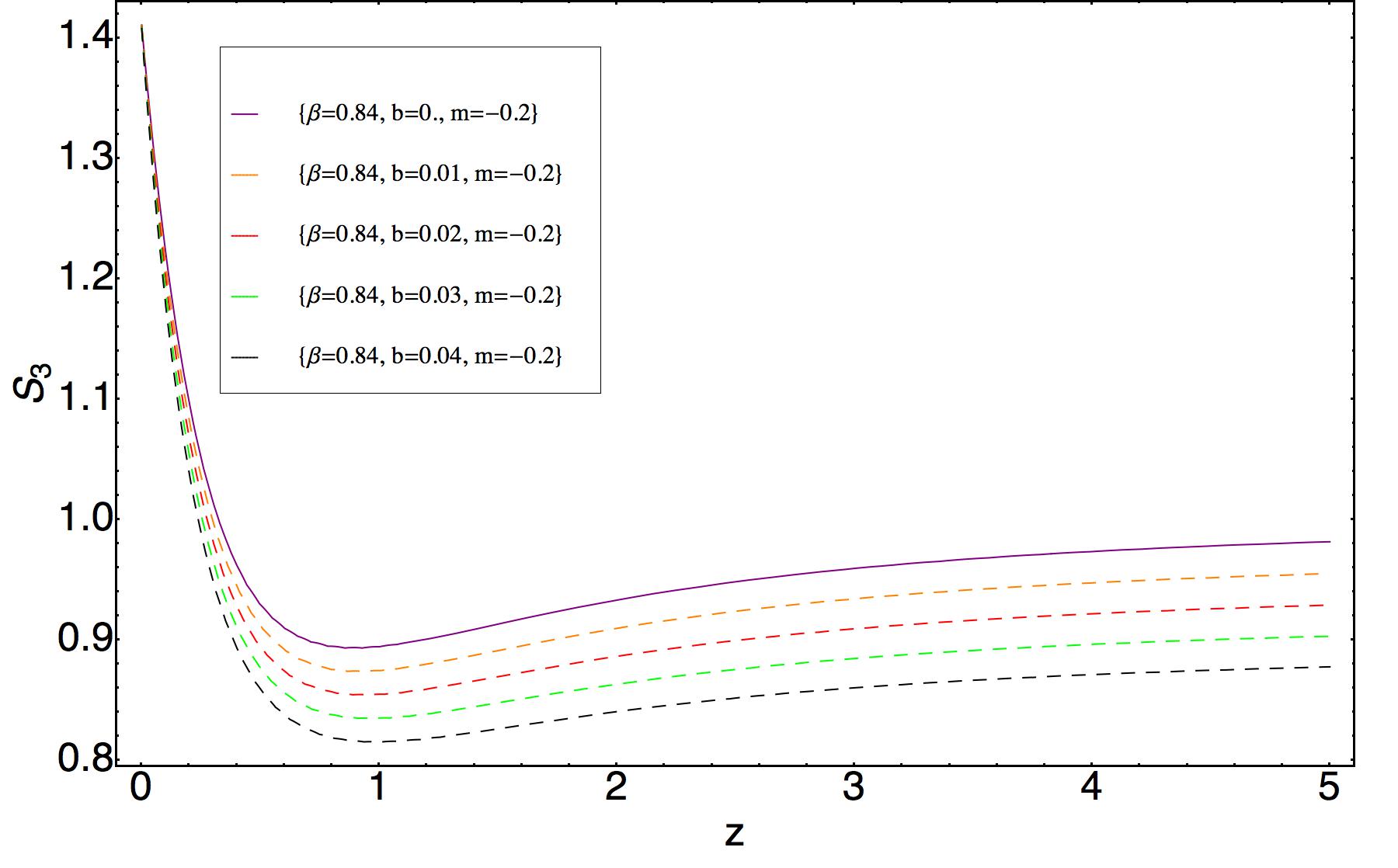}  \\
 \includegraphics[width=80 mm]{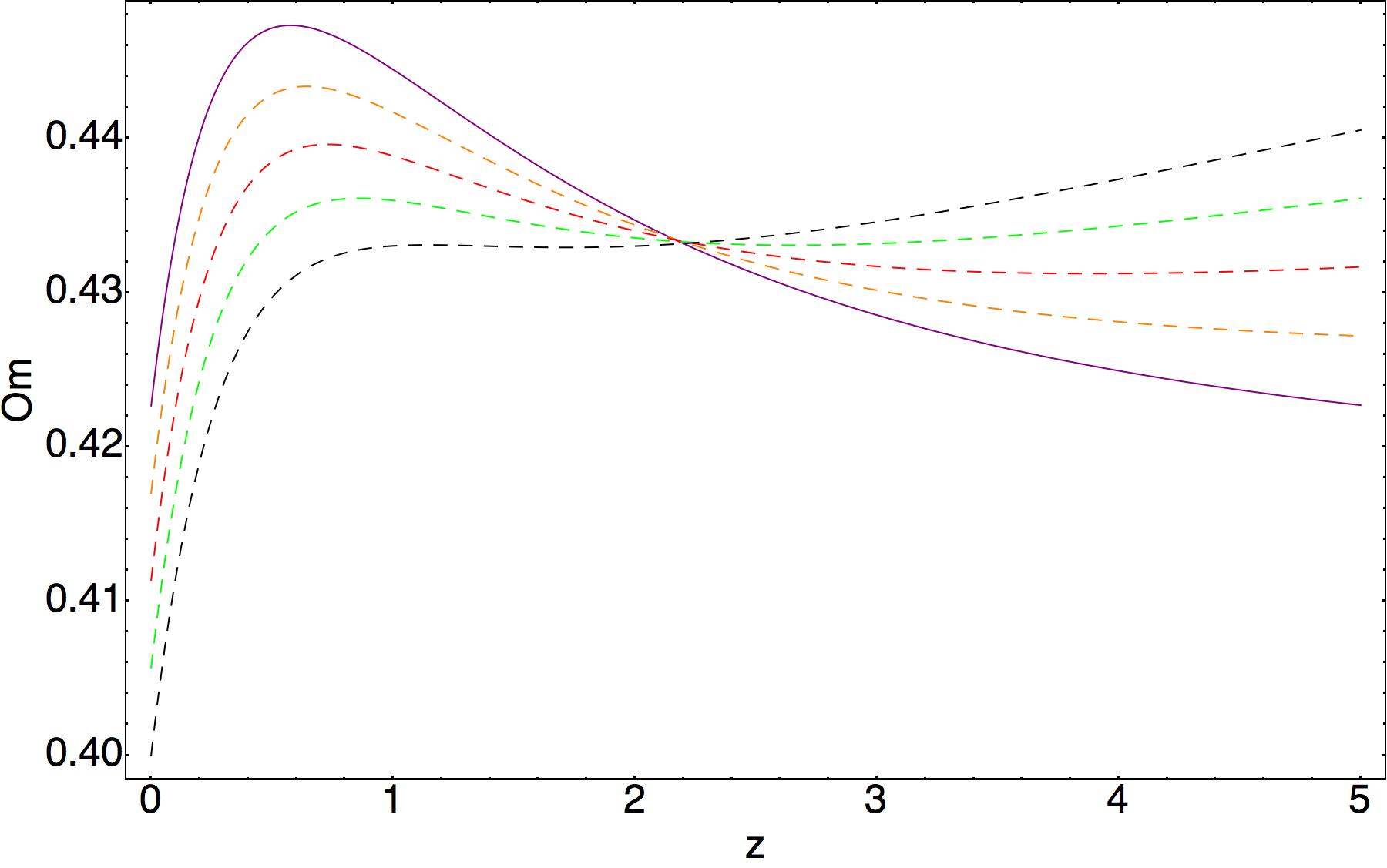}  &
\includegraphics[width=80 mm]{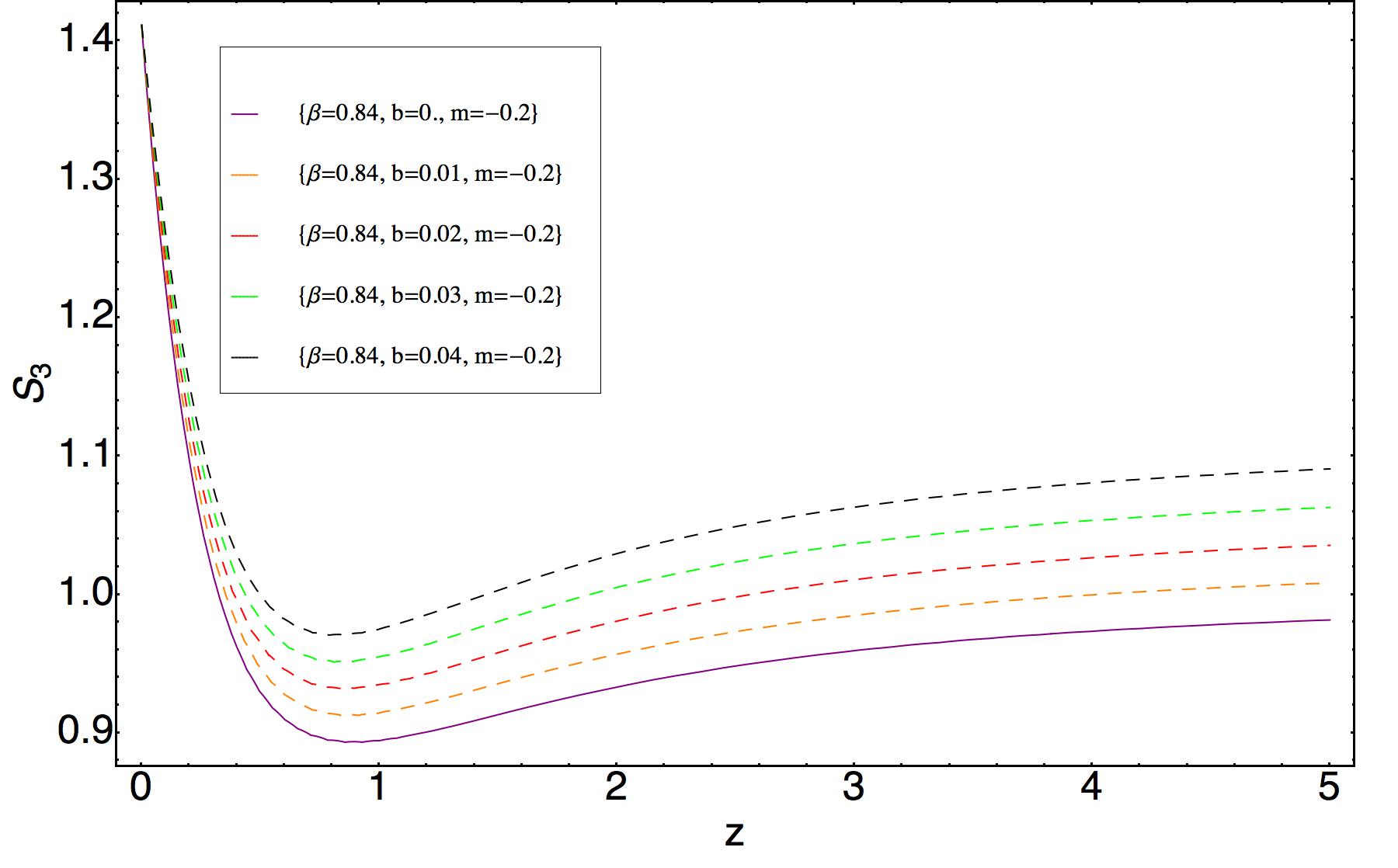}  \\
 \end{array}$
 \end{center}
\caption{Graphical behavior of $Om$ and $S_{3}$ parameters against redshift $z$. $m=0$ does correspond to usual ghost dark energy. Analysis presents cosmological model with interacting varying ghost dark energy Eq.~(\ref{eq:VGDE}). The interaction is given via Eq.~(\ref{eq:INT3}). Behavior of $Om$ parameter is according to the same values of the parameters of the model as for $S_{3}$ parameter.}
 \label{fig:Fig7}
\end{figure}

Various tools have been developed in order to distinguish dark energy models. During discussion of cosmography of suggested models, we have estimated present day value of $(r,s)$ and $(\omega_{de}^{\prime}, \omega_{de})$ parameters, which are one of the first parameters suggested to be studied in order to have appropriate discrimination of dark energy models. Since we have suggested new cosmological scenario, then it is reasonable to organize more study on them, particularly involving analysis allowing us to understand possible deviations of these models from other cosmological scenarios, particularly from $\Lambda$CDM model. In this section, for such purpose we will organize a detailed study of $Om$ parameter and statefinder hierarchy analysis for our models. $Om$ analysis suggests to study the following parameter
\begin{equation}
Om = \frac{x^{2}-1}{(1+z)^{3} - 1},
\end{equation}  
where $x = H/H_{0}$, $H$ it is the Hubble parameter and $H_{0}$ it is the value of the Hubble parameter at $z=0$. While statefinder hierarchy requires to calculate and study the following parameters 
\begin{equation}\label{eq:S3}
S^{(1)}_{3} = A_{3},
\end{equation}
\begin{equation}\label{eq:S4}
S^{(1)}_{4} = A_{4} + 3(1+q),
\end{equation}
\begin{equation}\label{eq:S5}
S^{(1)}_{5} = A_{5}  - 2 (4+ 3q)(1+q),
\end{equation}
etc., where $q$ it is the deceleration parameter, while $A_{n}$ does read as
\begin{equation}
A_{n} = \frac{a^{(n)}}{a H^{n}},
\end{equation} 
with 
\begin{equation}
a^{(n)} = \frac{d^{n}a}{dt^{n}}.
\end{equation}
Statefinder hierarchy for $\Lambda$CDM model during the cosmic expansion is equal to $1$. However, for the models with dynamical dark energy, dark matter and radiation $S^{(1)}_{n}$ is a varying quantity and $\Lambda$CDM model can be chosen as a reference frame to emphasize possible deviations. On the other hand, for $\Lambda$CDM model $Om = \Omega_{m0}$ i.e. in our case $Om = 0.3$ and it will be chosen as a reference frame for this analysis. In Fig.~(\ref{fig:Fig4}) we have presented behavior of $Om$ and $S_{3}$ parameters for non interacting model for different values of $m$ as it has been discussed in subsection~\ref{ssec:NINT}. We see, that $Om$ parameter is well above from $Om = 0.3$ line, moreover, an increasing of $m$~(started from an appropriate value of $m$) will provide only an increasing $Om$ parameter. For the set of parameters of the non interacting model giving the best fit of theoretical results to the distance modulus~(in our case the blue line in the left plot for $m=-0.2$) $Om$ parameter is a slowly increasing function for high redshifts, while for low redshifts it is a decreasing function. Especially, this decreasing behavior observed for $Om$ parameter will disappear with increasing of $m$. In the right plot of Fig.~(\ref{fig:Fig4}) we have presented redshift dependent behavior of $S_{3}$ parameter from statefinder hierarchy. It is a good indicator to distinguish the models, therefore we will concentrate our attention on it only. We see, that an increase of the parameter $m$ will increase the difference between suggested model and $\Lambda$CDM. $S_{3}$ parameter it is an increasing function and an increase of $m$ will increase its present day value. Fig.~(\ref{fig:Fig5}) presents redshift dependence of $Om$ and $S_{3}$ for the cosmological model where the interaction between the dark components is given by Eq.~(\ref{eq:INT1}). Left plot of Fig.~(\ref{fig:Fig5}) shows that consideration of the interaction of this particular type, will decrease $Om$ parameter according to $Om$ parameter for non interacting case, when we will increase the value of parameter $b$ describing the strength of the interaction. The same could be said about $S_{3}$ parameter presented in the right plot of Fig.~(\ref{fig:Fig5}). On the other hand, if we consider the sign changeable interaction given by Eq.~(\ref{eq:INT2}), then an increase of $b$ will decrease $Om$. However, for redshifts $z < 1.1$, the nature of $Om$ parameter will be changed making it an increasing function~(left plot of Fig.~(\ref{fig:Fig6})). Behavior of $S_{3}$ parameter presented on the right plot of Fig.~(\ref{fig:Fig6}) indicates that $S_{3}$ is a decreasing and an increasing function for low redshifts. Appropriate conclusion about impact of $b$ parameter on $S_{3}$ can be seen from Fig.~(\ref{fig:Fig6}). Plots presented in Fig.~(\ref{fig:Fig7}) correspond to the models considered in subsection~\ref{ssec:INT3}. We see that $Om$ analysis is more appropriate analysis to distinguish these interacting models from each other for different values of the parameter $b$, then $S_{3}$ parameter from the statefinder hierarchy~(for low redshifts). On the other hand, both parameters are good indicators to distinguish considered models from $\Lambda$CDM model.     

\section{\large{Discussion}}\label{sec:Discussion}
Among various dark energy models actively studied in Literature, in this paper we have worked out a cosmographic picture corresponding to the large scale universe containing a varying ghost dark energy. Considered varying ghost dark energy it is a phenomenological modification of the ghost dark energy of a specific form. Motivation behind such modification, we thought, is related to physics making a connection between the dark matter and the geometry~(expressed in terms of the Hubble parameter). Particularly, our phenomenological consideration is due to an existence of a possible non minimal coupling~(yet unknown) between the geometry and non relativistic matter. In Refs.~\cite{S0}~-~\cite{S1} various ways to generalise the simplest barotropic fluid EoS to inhomogenious fluids EoS has been suggested taking into account possible interplay between the pressure, the energy density and the Hubble parameter, among other possibilities. Particularly, suggested model of the dark energy belongs to the fluids which satisfies to more general form of the EoS given by $F(P_{de},\rho, H)$, where $\rho$ could be the energy density either of the effective fluid, or one of the components of the darkness, while $H$ it is the Hubble parameter and $P_{de}$ it is the pressure. Recently, other models of the varying ghost dark energy have been considered in Literature. Moreover, for considered models, it has been shown that in an appropriate radiation dominated expanding universe, which evolves to the large scale universe with suggested varying ghost dark energies, massless particle creation is possible. Such study concerning to suggested varying ghost dark energy of this paper we have left as a subject of another discussion elsewhere, particularly, when deep comparison of the theoretical results with the observation data will be performed and appropriate constraints on the parameters of the models will be obtained. On the other hand, our study in this paper showed us that non interacting model of the varying ghost dark energy for appropriate value of new parameter $m$ gives theoretical results well comparable to observational data. Non interacting model considered in this paper is the minimal model among considered models. Taking into account an active discussion on an interaction between the dark energy and nonrelativistic dark matter, we considered impact of various forms of interaction on cosmography. Particular type of interaction considered in this paper left unique trace in the behavior of appropriate cosmological parameters. Since there is an active discussion on sign changeable interactions in Literature, we have also our study of appropriate cosmological scenarios. Generally, $3$ forms of sign changeable interaction have been considered and discussed. On the other hand, in case of non interacting model and in case of interacting models of the varying ghost dark energy, from the redshift dependent profiles of the deceleration parameter we observed the phase transition from a decelerated expanding universe to the recent large scale universe. Moreover, present day values of $(r,s)$ and $(\omega_{de}^{\prime}, \omega_{de})$ parameters have been estimated completing the study of the models by $Om$ and statefinder hierarchy analysis. As was expected these two analysis are enough to have understanding about departures of our models from $\Lambda$CDM standard model. Moreover, these analyses are able to demonstrate how considered interactions affects on the departures of the interacting models from each other. Considered constraints on the cosmological parameters are due to PLANCK 2015 satellite experiment~\cite{PlanckCol}, which has been completed by the constraints on the parameters of the models due to the best fit of the theoretical results to the distance modulus. Considered modification of the ghost dark energy provides interesting departures. Another interesting question that could be studied within suggested cosmological scenarios it is the future of the recent universe. There are $4$ types of the singularity actively discussed in Literature that our universe can evolve eventually and the properties of the singularities in (phantom) dark energy universe were introduced at first time in Ref.~\cite{S3}. Assuming that the physics after the redshift $z=0$ will not be changed and the content of the universe still can be approximated as a two component effective fluid considered in this paper, we observed Type II singularity to be the characteristic singularity for the future for our models. Our conclusion is due to the behavior of the pressure $|P| \to \infty$ with finit values for the scale factor $a$ and $\rho$, when $z \to -1$~(according to the values of the parameters of the models giving relevant large scale universe). Of course, consideration of the interaction can change the type of the singularity and evan can remove it from the model, however, as showed our study, in considered cases removing of the future singularity due to $m$ and $b$ parameters will cause the phase transition to the accelerated expanding universe happen for $z<0$ redshifts - directly contradicting to the recent observational data. Therefore, our next goal is to organize an appropriate discussion providing more physics behind suggested modification of the ghost dark energy completing our study by a study of massless particle creation in an appropriate radiation dominated expanding universe. Moreover, when a closer look to these models due to deep constraints on the parameters will be obtained, we will study structure formation problem providing appropriate explanation of the behavior $Om$ parameter seen in case of sign changeable interaction. We hope to report on these studies very soon elsewhere. Generalization of considered phenomenological models can involve consideration of nonlinear interactions and viscosity. On the other hand, it will be interesting to relate the varying ghost dark energy of this paper with dark energy describing string landscape~~\cite{S4}.

\newpage

\end{document}